\documentclass[fleqn,usenatbib]{mnras}

\usepackage{newtxtext,newtxmath}
\usepackage[T1]{fontenc}
\usepackage{ae,aecompl}

\usepackage{graphicx}	% Including figure files
\usepackage{amsmath}	% Advanced maths commands
\usepackage{amssymb}	% Extra maths symbols
\usepackage{subfig}	

\newcommand{\kmsmpc}{\kms\;{\rm Mpc}^{-1}}

\newcommand{\HI}{\ion{H}{i}}

\newcommand{\hkpc}{h^{-1}{\rm kpc}}
\newcommand{\hmpc}{h^{-1}{\rm Mpc}}

\newcommand{\kms}{\;{\rm km}\,{\rm s}^{-1}}

\newcommand{\gad}{{\sc Gadget-3}}
\newcommand{\gizmo}{{\sc Gizmo}}
\newcommand{\mufasa}{{\sc Mufasa}}
\newcommand{\simba}{{\sc Simba}}

\newcommand{\fedd}{f_{\rm Edd}}
\newcommand{\mbh}{M_{\rm BH}}
\newcommand{\diff}{{\rm d}}
\title[Simba]{\simba: Cosmological Simulations with Black Hole Growth and Feedback}

\author[Dav\'e et al.]{
Romeel Dav\'e$^{1,2,3}$, 
Daniel Angl\'es-Alc\'azar$^{4}$, 
Desika Narayanan$^{5,6,7}$,
\newauthor Qi Li$^5$,
Mika H. Rafieferantsoa$^{2,3}$,
\& Sarah Appleby$^1$
\\
\\$^1$ Institute for Astronomy, Royal Observatory, Univ. of Edinburgh, Edinburgh EH9 3HJ, UK
\\$^2$ University of the Western Cape, Bellville, Cape Town 7535, South Africa
\\$^3$ South African Astronomical Observatories, Observatory, Cape Town 7925, South Africa
\\$^4$ Center for Computational Astrophysics, Flatiron Institute, 162 Fifth Avenue, New York, NY 10010, USA
\\$^5$ Department of Astronomy, University of Florida, 211 Bryant Space Sciences Center, Gainesville, FL, USA
\\$^6$ University of Florida Informatics Institute, 432 Newell Drive, CISE Bldg E251, Gainesville, FL, USA
\\$^7$ Cosmic Dawn Center at the Niels Bohr Institute, University of Copenhagen\
 and DTU-Space, Technical University of Denmark}

\date{Accepted XXX. Received YYY; in original form ZZZ}

\pubyear{2018}

\hypersetup{draft}

\begin{document}
\label{firstpage}
\pagerange{\pageref{firstpage}--\pageref{lastpage}}
\maketitle

\begin{abstract}
We introduce the \simba\ simulations, the next generation of the \mufasa\ cosmological galaxy formation simulations run with \gizmo's meshless finite mass hydrodynamics. \simba\ includes updates to \mufasa's sub-resolution star formation and feedback prescriptions, and introduces black hole growth via the torque-limited accretion model of \citet{Angles:2017} from cold gas and Bondi accretion from hot gas, along with black hole feedback via kinetic bipolar outflows and X-ray energy.  Ejection velocities are taken to be $\sim 10^3\kms$ at high Eddington ratios, increasing to $\sim 8000\kms$ at Eddington ratios below 2\%, with a constant momentum input of $20L/c$.  \simba\ further includes an on-the-fly dust production, growth, and destruction model.  Our \simba\ run with $(100\hmpc)^3$ and $1024^3$ gas elements reproduces numerous observables, including galaxy stellar mass functions at $z=0-6$, the stellar mass--star formation rate main sequence, \HI\ and H$_2$ fractions, the mass-metallicity relation at $z\approx0,2$, star-forming galaxy sizes, hot gas fractions in massive halos, and $z=0$ galaxy dust properties.  However, \simba\ also yields an insufficiently sharp truncation of the $z=0$ mass function, and too-large sizes for low-mass quenched galaxies.  We show that \simba's jet feedback is primarily responsible for quenching massive galaxies.
\end{abstract}


%%%%%%%%%%%%%%%%%%%%%%%%%%%%%%%%%%%%%%%%%%%%%%%%%%

%%%%%%%%%%%%%%%%% BODY OF PAPER %%%%%%%%%%%%%%%%%%

\section{Introduction}

The formation and evolution of galaxies is governed by a wide range of physical processes spanning from sub-parsec to giga-parsec scales.
These include the growth of large-scale structure, cooling and heating of astrophysical plasmas, and the formation of stars and
central black holes along with associated energy return processes collectively known as feedback.  As observations of galaxies improve at an ever-advancing pace, a key goal of modern astrophysics is to understand how such observations can be used to constrain the balance of underlying physical processes driving galaxy evolution, particularly in regards to feedback processes that are currently among its least well-understood aspects.

Modern galaxy formation models are generally built on the premise that feedback at the very lowest masses is dominated by photoionisation from the metagalactic ultraviolet background, feedback at scales above this but below $L^\star$ is  driven primarily by energy and momentum from young stars and supernovae, and feedback in massive galaxies predominantly owes to energetic release from accretion disks around supermassive central black holes~\citep[see][for a review]{Somerville:2015}.  While this framework has been broadly successful at reproducing many key observed characteristics of the galaxy population at a range of cosmic epochs, the physical understanding of most of these small-scale feedback processes remains coarse and heuristic~\citep[see][for a review]{Naab:2017}.  Galaxy formation models have now progressed to a point where numerous models can reproduce similar core sets of observations, but they often do so with significantly different underlying physical models and assumptions.  To discriminate among these physical drivers, it becomes important to develop modeling methodologies that are as well-motivated and realistic as possible, and to test such models against the widest possible suite of observations quantifying the stellar, gas, metal, and black hole properties of galaxies along with their surrounding gas.

Cosmological-scale simulations that model galaxy growth and feedback dynamically within evolving large-scale structure are an increasingly valuable tool for testing and constraining galaxy formation physics, owing both to rapidly advancing computational power and the commensurate ability to concurrently model a large range of scales and physical processes.  State of the art models now simultaneously predict the co-evolution of stars, interstellar media, black holes, and circum-galactic gas, enabling a holistic approach towards testing the input physics against observations across a wide range of scales, environments, and wavelengths. Modern cosmological-scale simulations such as Illustris~\citep{Vogelsberger:2014,Genel:2014}, Magneticum~\citep{Hirschmann:2014}, Horizon-AGN~\citep{Dubois:2014,Volonteri:2016,Kaviraj:2017}, Eagle~\citep{Schaye:2015}, MassiveBlack~\citep{Khandai:2015}, Blue Tides~\citep{Feng:2016}, Romulus~\citep{Tremmel:2017}, and Illustris-TNG~\citep{Springel:2018} have implemented ever-improving sub-grid models aimed at more successfully reproducing the stellar, gaseous, and black hole contents of galaxies in bulk, while numerous associated hierarchically-situated zoom simulations using more detailed input physics can examine the internal structural and dynamical properties of galaxies with increasing fidelity.

The \mufasa\ simulation project~\citep{Dave:2016} has added to the pantheon of such simulations, employing several novel approaches that distinguish it from others.  First, it utilises meshless finite mass (MFM) hydrodynamics as implemented in the \gizmo\ code~\citep{Hopkins:2015,Hopkins:2018}, which offers important accuracy advantages over Smoothed Particle Hydrodynamics (SPH), and owing to the mass conserving nature of its gas elements greater ease of analysis as compared to adaptive mesh refinement ({\sc Ramses}) and moving mesh ({\sc Arepo}) codes.  Second, rather than employing simple parameterisations or a cooling shutoff to drive galactic outflows, the kinetic mass outflow rate is taken directly from very high-resolution simulations from the Feedback in Realistic Environments~\citep[FIRE;][]{Hopkins:2014,HopkinsFIRE:2018,Muratov:2015} simulations, providing a synergy between ISM-resolving simulations of individual galaxies and cosmological-scale simulations of galaxy populations.  

An aspect where \mufasa\ was physically less well motivated than other state of the art simulations was in its depiction of black hole growth and AGN feedback.  \mufasa\ did not directly grow black holes and utilise the accretion energy for feedback to quench galaxies.  Instead, following \citet{Gabor:2015}, \mufasa\ implemented a heuristic ``quenching feedback'' in which diffuse gas within massive halos was prevented from cooling.  This energy was envisioned to be putatively supplied by an AGN, but \mufasa\ did not explicitly model black hole accretion and the interaction of its energy release with surrounding gas.  The halo mass scale above which quenching feedback was applied was taken from the best-fit analytic equilibrium model of \citet{Mitra:2015,Mitra:2017}, and evolved slowly upwards with redshift.  Despite its simplicity, this prescription displayed impressive successes at reproducing a red sequence and massive central galaxy properties in excellent accord with observations~\citep{Dave:2017b}, albeit with some non-trivial discrepancies such as over-quenching satellites particularly in the outskirts of large halos~\citep{Rafieferantsoa:2018}.  This demonstrated that a model based primarily on starvation of the central galaxy via ``radio mode" feedback~\citep{Croton:2006,Bower:2006} is able to quench the galaxy population in a hydrodynamic simulation in broad agreement with observations.  While this model represented an interesting numerical experiment, it would clearly be valuable to implement a more physically-motivated black hole growth and feedback model that retains and perhaps even extends the successes of \mufasa's more heuristic model.

This is the primary goal of the \simba\ project.  As the descendent of \mufasa, \simba\ marries two lines of investigation to achieve this.  First, it builds on the successful \mufasa\ model, including its representation of star formation-driven feedback and other modern features.  To this, \simba\ adds a novel and promising model for black hole growth: Torque-limited accretion~\citep{Hopkins:2011,Angles:2013,Angles:2015,Angles:2017}.  In this model, black hole growth is regulated by the ability for gas in the inner disk to lose angular momentum via disk instabilities.  \citet{Hopkins:2011} developed an analytic formalism for this, tested and calibrated using sub-pc scale numerical simulations, which yielded a formula that connects the infall rate of material onto the black hole accretion disk with properties of the inner galactic disk.  They showed that even at $\sim 1$~kpc resolution typical of cosmological simulations, their gravitational torque accretion formula provides a significantly better match to the measured accretion rate in their high-resolution simulations than employing the canonical Bondi accretion formula used in all other current cosmological black hole growth simulations. 

\citet{Angles:2013} explored the torque-limited accretion model via the post-processing of zoom simulations, and \citet{Angles:2015} extended this approach to cosmological simulations.  Their most significant result was that, unlike Bondi accretion, torque-limited accretion does not require the black hole to self-regulate its own growth.  In particular, torque-limited accretion naturally results in black holes growing along the observed galaxy-black hole scaling relations, even without any black hole feedback.  There is one free parameter in the model which represents the fraction of material entering the accretion disk that accretes onto the black hole; a plausible choice of $\sim 10$\% provided a good match to data, insensitive to the choice of the (uncertain) black hole seed mass. \citet{Angles:2017} extended these previous works to self-consistently incorporate torque-limited accretion into \gizmo, along with bipolar black hole winds, and demonstrated that the results obtained without feedback were reproduced in this case -- in particular, the inclusion of feedback self-consistently confirmed the primary result obtained in the post-processed case that black hole--galaxy scaling relations arise naturally without the black hole self-regulating its own growth. \simba\ builds on this work to employ the torque-limited black hole accretion model of \citet{Angles:2017} when accreting from cold or star-forming gas, in order to self-consistently grow black holes within galaxies during the simulation run.  The use of torque-limited black hole growth is unique among current cosmological simulations. \simba\ also includes Bondi accretion, but only from hot gas when present close to the black hole since it is the physically appropriate model in that case.

The second part of \simba's new black hole model involves a novel sub-grid prescription for active galactic nuclei (AGN) feedback.  AGN feedback connects flows coming off the black hole accretion disk to energy release on scales of tens or hundreds of kpc. To model this transfer of energy from small to large scales, \simba\ utilises  kinetic outflows with outflow parameters based on observed AGN feedback.  While there is still no well-defined theoretical consensus on the generation of black hole outflows and jets, recent observational progress has been rapid, showing that AGN can drive molecular and ionised outflows with velocities of $\sim 1000\kms$ or more~\citep{Sturm:2011,Greene:2012,Maiolino:2012,Liu:2013,Perna:2017a}, and jets at velocities up to $\sim 10^4\kms$ and more~\citep{Fabian:2012}.  Generally, high-velocity jets are observed to arise from early-type galaxies hosting massive black holes with low accretion rates relative to its Eddington rate ($\fedd\la$few percent), while lower-velocity outflows typically arise in systems with higher $\fedd$~\citep{Best:2012,Heckman:2014}. Extreme systems such as bright quasars often show both types of outflows. \simba's black hole outflows are parameterised to broadly follow such observed trends.  
\simba\ employs kinetic feedback (i.e. gas element kicks) for both feedback modes, with the kick velocity ranging from many hundreds of km/s in low-mass, fast-accreting black holes, up to many thousands of $\kms$ for slower-accreting black holes.

A key unique feature is that \simba's kinetic feedback is purely bipolar, with the ejection vector given by the angular momentum of the inner disk. This direction is relatively stable over galaxy dynamical timescales. To be conservative in minimising black hole feedback impact on the galaxy interstellar medium, we employ an opening angle of zero.  This is in contrast to other simulations that successfully reproduce massive galaxy properties using Bondi accretion, which employ either spherical thermal input~\cite[e.g. EAGLE;][]{Schaye:2015} or randomise the kinetic feedback's direction on short timescales~\citep[e.g. Illustris-TNG;][]{Weinberger:2017}.  {\color{black} Horizon-AGN employed Bondi accretion with a two-mode feedback scheme~\citep{Dubois:2012}, but still used spherical thermal feedback during the high-Eddington growth phase, as later also done in Illustris-TNG.  More detailed isolated elliptical simulations have also highlighted the importance of radiative mechanical feedback ~\citep{Gan:2014} that can reproduce observations of AGNs in ellipticals such as their duty cycle~\citep{Gan:2019}.}

The reason \simba\ is able to be successful with a genuinely bipolar model likely traces back to its accretion model:  Torque-limited accretion does not require self-regulation of black hole growth, whereas Bondi accretion requires quasi-spherical energy distribution close to the black hole in order to self-regulate its own growth.  In \simba's case, the energy input sphericalises at large distances, sufficient in massive halos to quench inflow into the galaxy by keeping the halo gas hot. In this way, \simba's accretion and feedback models work together to build a sub-grid description of AGN feedback that is more closely connected to observations of AGN winds and jets.

In addition to kinetic AGN feedback, \simba\ also includes X-ray feedback input into surrounding gas.  The importance of this feedback channel has been emphasised in zoom simulations by \citet{Choi:2012}, showing that it can potentially drive the quenching of massive galaxies.  We adapt this model to operate under the lower-resolution conditions present in \simba's cosmological-scales runs, and show that it plays a minor but non-trivial role in quenching the most massive galaxies. \simba\ is the first cosmological-volume simulation to include such X-ray feedback.

Another novel aspect of \simba\ is that it includes a model for on-the-fly dust production and destruction, broadly following \citet{McKinnon:2017}'s implementation into {\sc Arepo} where the dust is passively advected with the gas. We include dust production from Type II supernovae (SNe) and Asymptotic Giant Branch (AGB) stars, and further growth via condensation from metals, while destruction can occur from sputtering, consumption by star formation, or SNe shocks.  The fraction of metals locked into dust can be substantial, leading to significant changes in the predicted mass-metallicity relations.  In \mufasa, we found it necessary to reduce the SN yields arbitrarily by a factor of two in order to match the observed gas-phase mass-metallicity relation, but in \simba\ we can reproduce this as well or better without such arbitrary factors, since a substantial fraction of the metals ends up locked in dust.

In this paper, we describe the simulation methodology in \simba\ which, besides the new black hole model, also makes various other minor improvements to \mufasa\ (\S\ref{sec:sim}).  We then present a range of observational comparisons to predicted stellar, gas, and metal properties, analogous to a sampling of results presented for \mufasa\ in a series of recent papers~\citep{Dave:2016,Dave:2017a,Dave:2017b,Rafieferantsoa:2018}, paying particular attention to black hole and massive galaxy properties which represent the most direct test of \simba's new AGN feedback model (\S\ref{sec:results}).  We show that \simba\ reproduces many observations comparably well or better than \mufasa, but now with a more realistic and self-consistent description of black hole growth and feedback.   We then examine variants of \simba's AGN feedback model in order to isolate the impact of its various components (\S\ref{sec:variants}), showing that the high-velocity jet feedback is crucial for producing a quenched massive galaxy population.  Finally, we summarize our results in \S\ref{sec:summary}.

\section{Simulation Methodology}\label{sec:sim}

\subsection{Code and input physics}\label{sec:code}

The \simba\ simulations utilise much of the framework of the \mufasa\ simulations as described in \citet{Dave:2016}, but there are a number of updates and additions based on recent theoretical and observations results, in addition to the major change of modeling black hole growth and feedback as well as dust.  Here we recap the main features of the model, and then describe in more detail the new aspects of \simba.

\simba\ utilises a forked version of the \gizmo\ cosmological gravity plus hydrodynamics solver~\citep{Hopkins:2015,Hopkins:2018}, in its Meshless Finite Mass (MFM) version.  This code, based on \gad~\citep{Springel:2005}, evolves dark matter and gas elements together including gravity and pressure forces, handling shocks via a Riemann solver with no artificial viscosity.  It performs very well in numerous standard hydrodynamics tests including strong shocks, rotating disks, and cold blobs moving through hot media~\citep{Hopkins:2015}.  It also preserves the mass within each fluid element during the evolution, thereby enabling detailed tracking of gas flows.  It thus marries the advantages of a particle-based code such as adaptivity in space and time, with the hydrodynamics accuracy of a Riemann solved-based mesh code, without the imposition of a Cartesian mesh that can cause difficulties with Galilean invariance and rotating shear flows.

Radiative cooling and photoionisation heating are modeled using the {\sc Grackle-3.1} library~\citep{Smith:2017}, including metal cooling and non-equilibrium evolution of primordial elements.  This is an updated version of {\sc Grackle-2.1} used in \mufasa, offering two advantages: First, the adiabatic and radiative terms are evolved together during the cooling sub-timestep, unlike the previous operator-split approach where first the system was evolved adiabatically over the full timestep and then cooling was applied; this results in more accurate and stable thermal evolution particularly in the stiff regions of the cooling curve.  Second, it includes self-shielding self-consistently during the simulation run, based on the \citet{Rahmati:2013} prescription in which the ionising background strength is attenuated depending (primarily) on gas density.  A spatially-uniform ionising background is assumed as specified by \citet{Haardt:2012}, modified to account for self-shielding (A. Emerick, priv. comm.).  These changes do not make a noticeable difference to the resulting galaxy population, but they do improve the accuracy of the baryonic thermal evolution which may be particularly important within circum-galactic gas.  Furthermore, computing the neutral hydrogen content of gas particles is now done self-consistently within the code rather than via a post-processed application of self-shielding~\citep{Dave:2017a}.

As in \mufasa, we use an H$_2$-based star formation rate, where the H$_2$ fraction is computed based on the sub-grid model of \citet{Krumholz:2011} based on the metallicity and local column density, with minor modifications as described in \citet{Dave:2016} to account for variations in numerical resolution.  The star formation rate is given by the H$_2$ density divided by the dynamical time: SFR$=\epsilon_*\rho_{H2}/t_{\rm dyn}$, where we use $\epsilon_*=0.02$~\citep{Kennicutt:1998}.
The chemical enrichment model tracks eleven elements (H,He,C,N,O,Ne,Mg,Si,S,Ca,Fe) during the simulation, with enrichment tracked from Type II supernovae (SNe), Type Ia SNe, and Asymptotic Giant Branch (AGB) stars. {\color{black} The yield tables employed are the same as in \mufasa, namely \citet{Nomoto:2006} for SNII yields, \citet{Iwamoto:1999} for SNIa yields, and AGB star enrichment following \citet{Oppenheimer:2006}.  However,} we no longer apply an arbitrary reduction of yields by a factor of 2 that was previously needed to match the mass-metallicity relation, and instead lock individual metals into dust; we detail the dust model implementation in \S\ref{sec:dust}.  Type Ia SNe and AGB wind heating are also included as in \mufasa, along with ISM pressurisation at a minimum level as required to resolve the Jeans mass in star-forming gas as described in \citet{Dave:2016}.

\begin{figure}
	\includegraphics[width=\columnwidth]{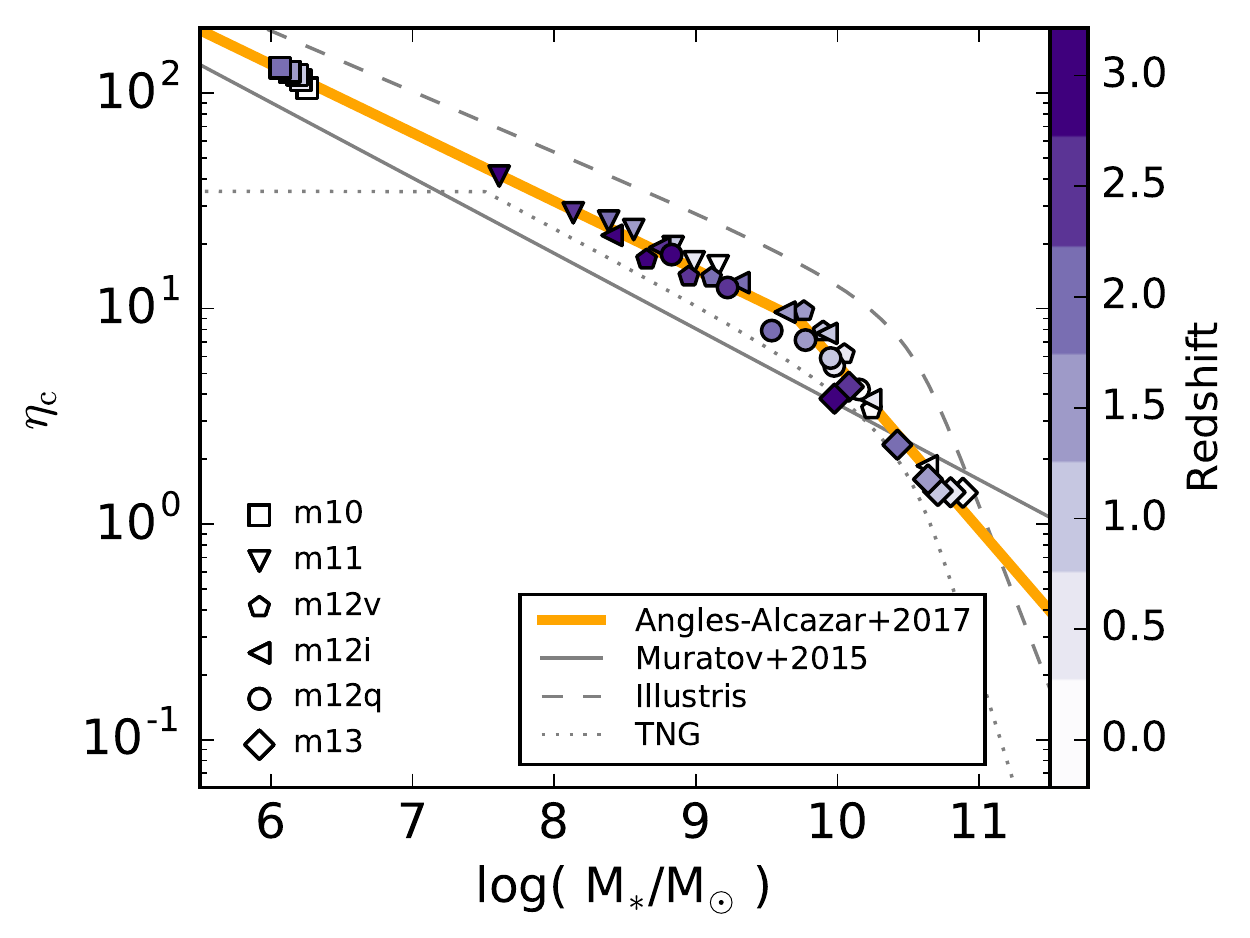}
	\vskip-0.1in
    \caption{Mass loading factor $\eta$ versus stellar mass $M_*$ from a suite of FIRE simulations analysed via particle tracking in \citet{Angles:2017b}.  The points show values measured at various redshifts, while the orange line is the best-fit relation.
    {\color{black} The gray solid line shows the \citet{Muratov:2015} scaling and the dashed and dotted gray lines show the mass loading factors used in the Illustris and Illustris-TNG simulations ($\eta(M_{\rm halo})$ fitting functions from \citealt{Pillepich:2018}, converted to $\eta(M_*)$ using the $M_*$--$M_{\rm halo}$ relation of \citealt{Moster:2013}).}}
    \label{fig:eta_mstar}
\end{figure}

The model for star formation-driven galactic winds closely follows that in \mufasa; we continue to use decoupled two-phase winds, with 30\% of wind particles ejected ``hot" i.e. with a temperature set by the supernova energy minus the wind kinetic energy.  However, we make a significant update to the mass loading factor scaling with stellar mass. \mufasa\ used the scalings taken from \citet{Muratov:2015}, who computed the outflow rates based on mass advection across a boundary at one-quarter of the virial radius in the FIRE zoom simulations. \citet{Angles:2017b} used similar FIRE simulations, but instead tracked individual particles in order to quantify the mass outflow rates out of the star-forming region, thus providing a more direct measurement of the amount of material leaving the ISM.  This yields a cumulative mass loading factor versus stellar mass as shown in Figure~\ref{fig:eta_mstar}, which is well fit by a broken power law at $M_0=5.2\times 10^9 M_\odot$:
\begin{equation}
\eta(M_*) \propto 
\begin{cases}
    9\Big(\frac{M_*}{M_0}\Big)^{-0.317},& \text{if } M_*<M_0\\
    9\Big(\frac{M_*}{M_0}\Big)^{-0.761},& \text{if } M_*>M_0\\
\end{cases}
\end{equation}
and is independent of redshift.  This has a similar slope to \citet{Muratov:2015} below $M_0$ but with roughly double the amplitude, and is much steeper above $M_0$.  
{\color{black} It is also similar to the assumed mass loading factor in Illustris-TNG~\citep{Pillepich:2018}.  Similar to TNG, \simba\ employs a flat $\eta(M_*)$ below an $M_*$ corresponding to 16 star particles ($M_*\leq 2.9\times 10^8 M_\odot$ for the $100\hmpc$ run), otherwise poorly-resolved forming galaxies are unable to grow owing to excessive feedback.}
As in \mufasa, we further apply a reduction in $\eta$ at high redshifts, in order to allow for early galaxy growth in poorly resolved situations.  \mufasa\ was found to underproduce $z>6$ galaxies, and hence we strengthen the suppression factor at $z>3$ from $(a/0.25)$ to $(a/0.25)^{f_a}$, where $a$ is the expansion factor.  We tune the value of $f_a$ based on the resolution of the simulation, since the origin of the lack of early galaxy formation owes to poor resolution.  Testing has shown that we obtain converged results that match $z\ga 6$ observations (shown later) if we use $f_a=2$ at our largest ($100\hmpc$) volume's resolution, $f_a=1.5$ at $8\times$ higher mass resolution, and so on.  Fortunately, because galaxy growth is very rapid at high redshifts, this choice makes little difference to galaxy predictions at $z\ga 3$ over most of cosmic time.  Note that the FIRE simulations do not make strong predictions for $\eta(M_*)$ at $z\ga 3$ owing to the limited dynamic range covered by their zooms at early epochs, so this choice is not in obvious conflict with using FIRE scalings at lower redshifts.   However, the ad hoc nature of this correction means that results for galaxy stellar growth at high redshifts from \simba\ should be considered as tuned rather than predictive.

{\color{black} A new feature in \simba\ is metal-loaded winds.  When a wind particle is launched, it extracts some metals from nearby particles to represent the local enrichment by the supernovae driving the wind.  The metallicity added to the wind particle is given by 
\begin{equation}
    dZ = f_{\rm SNII} y_{\rm SNII}(Z) / {\rm MAX}(\eta,1)
\end{equation}
where $f_{\rm SNII}=0.18$ is the stellar mass fraction lost to supernova (assumed to be instantaneous in \simba), $y_{\rm SNII}(Z)$ is the metal-dependent Type II SN yield for each species, and $\eta$ is the mass loading factor.  This amount is subtracted from nearby gas in a kernel-weighted manner.  If there is not enough metals nearby (as can happen early on), then $dZ$ is 
reduced appropriately.  In all circumstances, the total metal mass is conserved.  The metal loading factor (i.e. the ejected metallicity relative to surrounding gas) can be a factor of two or larger when the ISM has a metallicity $\ll y_{\rm SNII}$, but is more typically around 10-20\%, in broad agreement with metal loading estimates from zoom simulations~\citep{Muratov:2017,Christensen:2018}.}

\simba\ continues to use the wind velocity scalings from \citet{Muratov:2015} as in \mufasa, since the scaling follows the expected quasi-linear scaling of wind speed with escape velocity as observed~\citep[e.g.][]{Martin:2005}, and also because a full analysis of the velocity distribution of outflowing gas is not available from FIRE. In \mufasa, the amplitude was taken as a tunable parameter, and set to $2v_{\rm circ}$ at $v_{\rm circ}=200\kms$. Owing to the increase in the mass loading factor from low-mass galaxies, we find that a somewhat lower value of the wind velocity is required to compensate for this, so in \simba\ we reduce the normalisation to 1.6:
\begin{equation}
v_w = 1.6 \Big(\frac{v_{\rm circ}}{200 \kms}\Big)^{0.12} v_{\rm circ} + \Delta v(0.25R_{\rm vir})
\end{equation}
where $\Delta v(0.25R_{\rm vir})$ is the velocity corresponding to the potential difference between the launch point and one-quarter of the virial radius \citep[see][]{Dave:2016}.  A related new aspect in \simba\ is that we limit the wind kinetic energy to the available supernova energy by attenuating the wind speed when needed, although this only has noticeable effect in small galaxies at very early epochs.

\simba\ uses an on-the-fly approximate friends-of-friends (FOF) finder applied to stars and dense gas as described in \citet{Dave:2016} in order to compute the galaxy properties such as $M_*$, and as in \mufasa\ obtains $v_{\rm circ}$ using a scaling based on the baryonic Tully-Fisher relation.  Besides algorithmic improvements to improve parallel performance, the only change to this is that the FOF finder now also groups black holes into galaxies.

%A new feature of star formation driven winds is that we attenuate the wind speed at low redshifts.  This is motivated by detailed tracking of winds that show that, at $z\la 1$, outflows are significantly weaker once an ordered thin disk forms~\citep{Angles:2017}.  \citet{Faucher:2018} developed an analytic formalism to describe this, showing that the bursty star formation required to drive winds is only present at high redshifts or when the gas fraction is high.  We follow this formalism by including a criterion that the wind speed is only at its full value when the circular velocity is lower than $v_{\rm crit}=46 (f_{\rm gas}/0.2) \kms$, or $z>1.3$.  For $z<1.3$, we reduce the wind velocity by the ratio $v_{\rm crit}/v_{\rm circ}$.  We use the on-the-fly friends-of-friends algorithm to compute $v_{\rm circ}$ and $f_{\rm gas}$ as described in \citet{Dave:2016}.  This results in lower wind speeds in sizable galaxies at low redshifts, which is more consistent with the results from FIRE zooms.

\subsection{Black hole growth}

The most significant change in \simba\ relative to \mufasa\ is that black holes are seeded and grown during the simulation, and the accretion energy is used to drive feedback that serves to quench galaxies.  In this section we describe \simba's two-mode accretion model. The first mode closely follows the torque-limited accretion model presented in \citet{Angles:2017}, and we refer the reader there for full details.  The second mode uses Bondi accretion, but solely from the hot gas component.  In future work (Angl\'es-Alc\'azar et al., in prep.) we will show that this contribution from Bondi accretion is sub-dominant for all but the highest mass black holes.

\subsubsection{Torque-limited accretion from cold gas}

We model the gas inflow rate $\dot{M}_{\rm Torque}$ driven by disk gravitational instabilities from galactic scales down to the accretion disk surrounding the central black hole following \citet{Hopkins:2011}:

\begin{equation}\label{eq:torque} 
\begin{split}
\dot{M}_{\rm Torque} \; \approx \; \epsilon_{\rm T} \, f_{\rm d}^{5/2} \times 
\left ( \frac{M_{\rm BH}}{10^{8}\,{\rm M_{\odot}}} \right )^{1/6} 
\left ( \frac{M_{\rm enc}(R_{0})}{10^{9}\,{\rm M_{\odot}}} \right ) \\
\times \left ( \frac{R_{0}}{100\,{\rm pc}} \right )^{-3/2}  \left (1 +
\frac{f_{0}}{f_{\rm gas}} \right )^{-1} \, {\rm M_{\odot}\,yr^{-1}},
\end{split}
\end{equation}  
where $f_{\rm d}$ is the disk mass fraction (including both stars and gas), $M_{\rm enc}(R_{0})$ is the total gas+stellar mass, $f_{\rm gas}$ is the gas mass fraction in the disk component, $f_{0} \approx 0.31 \, f_{\rm d}^{2} \, (M_{\rm d}(R_{0})/10^{9}{\rm M_{\odot}})^{-1/3}$, and all quantities are evaluated within a distance $R_{0}$ of each black hole enclosing the nearest 256 gas elements, with an upper limit $R_{0} \leq 2\,\hkpc$ (comoving) imposed throughout the simulation.
Evaluating this equation for $\dot{M}_{\rm Torque}$ requires separating the spheroidal and disk components within $R_{0}$, which we do by means of the kinematic decomposition implemented in \citet{Angles:2017}; {\color{black} see the Appendix of \citet{Angles:2015} for further tests of this. }
Unlike previous work, we evaluate torque-limited accretion only for the cold ($T<10^{5}$~K) gas within the black hole kernel, since it relies on instabilities in a cold gaseous disk to drive mass inflow.  {\color{black} We consider all ISM gas to be in the cold phase, where ISM gas is taken to be gas that has been artificially pressurised in order to resolve the Jeans mass as described in \citet{Dave:2016}.  In our $100\hmpc$ run this corresponds to gas above a hydrogen number density of $n_H>0.13$~cm$^{-3}$ and within 0.5~dex of the pressurized ISM temperature floor.}

The normalization factor  $\epsilon_{\rm T} \equiv \epsilon_{\rm m} \times \alpha_{\rm T}$ encapsulates processes that affect the radial transport of gas on unresolved scales (e.g. nuclear star formation and stellar feedback and mass loss in winds from the accretion disk), where $\alpha_{\rm T} = 5$ is the original normalization of $\dot{M}_{\rm Torque}$ in \citet{Hopkins:2011} and we set $\epsilon_{\rm m} = 0.1$ to match the normalization of the local $M_{\rm BH}$--$M_\star$ relation as in \citet{Angles:2017}.  One can view $\alpha_{\rm T}$
as corresponding to an efficiency of transport of material from the inner galactic disk onto the black hole accretion disk, and $\epsilon_m$ as the efficiency of transport from the accretion disk onto the black hole, for which 10\% is a canonical value. However, $\alpha_{\rm T}$ is itself fairly uncertain, and in the end the meaningful subgrid parameter is only the combination $\epsilon_{\rm T}$.

\subsubsection{Bondi accretion from hot gas}

%Torque-limited accretion is only applied to star-forming (cold) gas, since it relies on disk instabilities to drive mass inflow.  
Hot gas can also accrete onto black holes, but in this case Bondi accretion is a more appropriate physical mechanism since the hot gas is typically more spherically distributed.  Thus we also account for Bondi accretion, but only from non-ISM gas with a temperature of $T>10^{5}$~K.

Bondi-Hoyle-Lyttleton accretion is computed via the standard \citet{Bondi:1952} formula:
\begin{equation}
\dot{M}_{\rm Bondi} = \epsilon_m \frac{4\pi G^2 \mbh^2 \rho}{(v^2+c_s^2)^{3/2}}
\end{equation}
where $\rho$ is the mean density of hot ($T>10^{5}$~K) gas computed within the black hole accretion kernel, $c_s$ is the kernel-averaged sound speed of the hot gas, and $v$ is the kernel-averaged velocity of the hot gas relative to the black hole.  In practice, we neglect the gas relative velocity and set $v=0$, since the dynamics of the black hole particle is controlled by the repositioning algorithm (\S\ref{sec:seeds}).     
We do not include any boost factor, since hot gas is likely distributed quite smoothly over the size of the black hole kernel. For consistency with the gravitational torque rate, we suppress Bondi accretion by the same efficiency $\epsilon_{\rm m} = 0.1$.
%The total accretion rate for a given black hole is then $\dot{M}_{\rm torque}+\dot{M}_{\rm Bondi}$.  

%\subsubsection{Variability and the Eddington limit}

%A new aspect relative to \citet{Angles:2017} is that we introduce an explicit model for black hole accretion variability. \citet{Kauffmann:2009} found that observed AGN variability can be described by a lognormal distribution with $\sigma=0.4$~dex.  Assuming that this variability reflects stochasticity in the accretion rate on short timescales that we cannot resolve directly, we multiply the above-computed accretion rate by a random multiplicative factor drawn from a distribution given by lognormal with $\sigma=0.4$~dex.  We correct for the fact that the mean value of a lognormal distribution is $e^{(\sigma\ln 10)^2/2}=1.53$ in our case, so that the mean accretion rate is unchanged from the torque-limited accretion formula.  Because our feedback model depends on the instantaneous accretion rate, such instantaneous variations can impact AGN feedback even though it does not change the overall average growth rate of black holes.

\subsubsection{Numerical implementation}
The total accretion rate for a given black hole is then 
\begin{equation}\label{eq:mdot} 
\dot{M}_{\rm BH} = (1 - \eta) \times \, (\dot{M}_{\rm Torque} + \dot{M}_{\rm Bondi}),
\end{equation}
where we adopt a constant radiative efficiency $\eta = 0.1$  \citep[e.g.][]{YuTremaine:2002}.  
We apply an overall limit to the accretion rate, based on the given black hole's Eddington accretion rate.  For  torque-limited accretion, we apply a limit of 3 times the Eddington rate, based on the idea that non-spherical accretion can potentially exceed Eddington as occasionally observed particularly at higher redshifts~\citep{Martinez:2018} and consistent with recent accretion disk simulations \citep[e.g.][]{Jiang:2014}.  For Bondi accretion, we apply a strict Eddington limit, since this is intended to represent quasi-spherical accretion from hot gas where the Eddington limit is directly applicable.

Numerically, we follow \citet{SpringelBH:2005} and track separately the physical black hole mass, which grows continuously according to equation~\ref{eq:mdot}, and the dynamical black hole particle mass, which grows by stochastically accreting a fraction $f_{\rm m}$ of the mass of gas particles within $R_0$ with a probability that statistically satisfies mass conservation and the desired mass outflow rate in AGN winds (see below).
A time step limiter is imposed on black hole particles such that black holes do not grow by more than 0.1\,\% of their current mass in a single time step.

\subsection{Black hole feedback}\label{sec:bh}

We incorporate a kinetic subgrid model for black hole feedback, along with X-ray energy feedback.  The motivation for the kinetic feedback model comes from the observed dichotomy in black hole growth modes that is reflected in their outflow characteristics~\citep[e.g.][]{Heckman:2014}:  A ``radiative mode" at high Eddington ratios ($\fedd \equiv \dot{M}_{\rm BH}/\dot{M}_{\rm Edd} \ga$ few percent), in which AGN are seen to drive multi-phase winds at velocities of $\sim 1000 \kms$ that include molecular and warm ionised gas~\citep[e.g.][]{Sturm:2011,Maiolino:2012,Perna:2017a}; and a ``jet mode" at low Eddington ratios, where AGN mostly drive hot gas in collimated jets at high velocities (of order $\sim 10^4\kms$) that in some circumstances are seen to inflate super-virial temperature bubbles in surrounding hot gas~\citep[e.g.][]{McNamara:2007,Fabian:2012}.  This dichotomy also appears in radio jet activity (``high excitation" vs. ``low excitation" radio galaxies) above and below roughly a percent of the Eddington rate~\citep{Best:2012}, with the former tending to be found in lower-mass, bluer host galaxies and the latter in massive early-types.  \simba's AGN feedback model is designed to directly mimic the energy injection into large-scale surrounding gas from these two modes, using purely bipolar feedback with velocities and temperatures taken as much as possible from AGN outflow observations.  We also include X-ray heating from black holes broadly following the model introduced by \citet{Choi:2012}, modified to operate at the lower resolution of \simba's cosmological simulations.  We now describe these subgrid models in more detail.

\subsubsection{Kinetic feedback}

For the high-$\fedd$ mode outflows, we choose an outflow velocity based on ionised gas linewidth observations of X-ray detected AGN from SDSS by \citet[see Fig. 8 of][]{Perna:2017a}, which we parameterise in terms of the black hole mass $\mbh$ (in $M_\odot$) as
\begin{equation}
v_{\rm w,EL} = 500+500(\log{\mbh}-6)/3\;\kms;
\label{eq:vradiative}
\end{equation} 
we refer to these as radiative AGN winds.  The gas is ejected without modifying its temperature, meaning that it is ejected at the ISM temperature given by our ISM pressurisation model. This is generally consistent with observations suggesting typical electron temperature of $\sim 10^4$K for the ionised gas outflows~\citep{Perna:2017b}.  This is broadly similar to the AGN feedback model implemented into \gizmo\ by \citet{Angles:2017}, except here with a variable outflow velocity.

If the Eddington ratio is below $\fedd<0.2$, we begin to transition to a jet mode with a velocity that becomes increasingly strong as $\fedd$ drops, as follows:
\begin{equation}
v_{\rm w,jet} = v_{\rm w,EL} + 7000 \log{(0.2/\fedd)}\;\kms,
\label{eq:vjet}
\end{equation}
with a cap to the velocity increase at $7000\kms$.  In this way, full speed jets are achieved only once $\fedd\la 0.02$.  To trigger jet mode, we also include an additional criterion requiring $\mbh>M_{\rm BH,lim}$, motivated by observations showing that radio jets only arise in galaxies with velocity dispersions corresponding to black holes with $\mbh\ga 10^8M_\odot$~\citep{Barisic:2017}.  To be conservative we choose $M_{\rm BH,lim}$ lower than this, namely $M_{\rm BH,lim}=10^{7.5}M_\odot$.  Physically, this mass limit is implemented in order to prevent small black holes that temporarily have low accretion rates from driving high-powered jets.

Based on observations of AGN outflows and the inferred momentum and energy input \citep[e.g.][]{Fiore:2017,Ishibashi:2018}, we set the amount of material ejected in the AGN winds in order to obtain a momentum input of $ \dot{P}_{\rm out} = 20\,L/c$, where $L = \eta \, \dot{M}_{\rm BH} \, c^2$ is the bolometric luminosity of the AGN, $\eta=0.1$, and $c$ is the speed of light.  This value is kept constant for both modes, resulting in the mass loading factor in AGN winds scaling inversely with the outflow velocity.  
For our parameter choices, a black hole with $M_{\rm BH} = 10^9\,M_\odot$ in the high-$\fedd$ mode injects outflows with $v_{\rm w,EL} = 1000\,\kms$, mass loading $\dot{M}_{\rm out,EL} / \dot{M}_{\rm BH} \approx 600$, and kinetic energy efficiency $\dot{E}_{\rm kin,EL} \approx 0.03\,L$, while in the low-$\fedd$ mode at full jet speed reaches $v_{\rm w,jet}=8000\kms$, $\dot{M}_{\rm out,jet} / \dot{M}_{\rm BH} \approx 75$, and $\dot{E}_{\rm kin,jet} \approx 0.3\,L$.

Particles are selected to be ejected randomly from within the black hole accretion kernel, with probability 
\begin{equation}\label{eq:stoch2} 
p_{j} \; = \; \frac{ 1 - f_{\rm m}}{f_{\rm m}} \times \frac{w_{j}}{m_j} \times \dot{M}_{\rm BH} \, \Delta t,  
\end{equation}
where $w_{j}$ is a kernel weight ($\Sigma_j \, w_j = 1$) and $f_{\rm m}$ is the fraction of  mass accreted by the black hole. 
The desired mass loading factor relative to the black hole accretion rate ($\dot{M}_{\rm out} / \dot{M}_{\rm BH} = (1 - f_{\rm m}) / f_{\rm m}$) is achieved by setting $f_{\rm m}$ such that:
\begin{equation}\label{eq:facc}
\frac{\dot{P}_{\rm out}}{L/c}  \, = \, 20 \, = \, \frac{v_{\rm w}}{\eta \,c} \, \left ( \frac{1 - f_{\rm m}}{f_{\rm m}} \right ).
\end{equation}

All outflows are ejected in a purely bipolar fashion.  That is, we eject gas elements in a direction $\pm$parallel to the angular momentum vector of the inner disk that we use to compute the black hole accretion (typically, the 256 nearest gas particle neighbours to the black hole).  We assume zero opening angle for all winds; this is probably conservative for the radiative mode winds, as the opening angles are likely to be wider, but for the jet mode it is a good approximation to observed highly collimated jets.  Even in the case of initially spherical radiative winds, it is likely that there is substantial collimation from the inner galaxy disk on scales that we cannot resolve in our cosmological runs, so the assumption of collimated winds is likely to be closer to correct.  Since the wind particles are launched from their current location, this results in a collimated outflow with a small but finite extent ($\la 1$~kpc).  We note that the outflow direction can precess owing to variations in the inner disk, but is in practice typically stable over tens to hundreds of Myr.  Hence any effect of ``sphericalising" the jet energy input on super-galactic scales is done self-consistently via the hydrodynamic interactions of the outflows with ambient gas at larger scales.

Since jets are observed to carry very hot gas, we raise the temperature of jet mode (only) outflows to the virial temperature of the halo, specifically $T_{\rm vir}=9.52\times 10^7 (M_{\rm halo}/10^{15}M_\odot)^{1/3}$~K~\citep{Voit:2005}.  This choice is motivated by observations showing that jets contain mostly synchrotron-emitting plasma, and eventually thermalise their energy into surrounding hot gas at around $T_{\rm vir}$~\citep{Fabian:2012}.  The extra energy input required for this is typically less than a few percent of the jet kinetic energy, so it does not figure significantly into the overall energy budget.

We apply a short hydrodynamic and radiative cooling decoupling time of $10^{-4}t_H$ to the outflowing wind gas elements, where $t_H$ is the Hubble time at launch.  This is in order to avoid further entrainment within the unresolved ISM close to the black hole, since the mass loading is accounted for from the assumption of constant momentum input of $20L/c$.  This also avoids some numerical inaccuracies from high Mach number shocks in very dense gas.  We note that for the jet mode, this can result in a decoupled distance of up to tens of kpc at the present epoch.  Hence the jet energy begins to be deposited at a distance comparable to the extent of observed radio lobes.  \gizmo\ employs a \citet{Durier:2012} timestep limiter in order to ensure proper interactions of the high-speed winds and their surrounding gas as they recouple.  

Our model has similarities to the two-mode thermal and kinetic AGN feedback model employed in Illustris-TNG~\citep{Weinberger:2017}.  The main differences are as follows: (i) Illustris-TNG uses Bondi accretion rather than torque-limited accretion for cold, rotationally supported gas.
(ii) Illustris-TNG uses spherical thermal feedback at high $\fedd$ rather than kinetic feedback.  This may owe to the fact that torque-limited accretion does not require self-regulation, while Bondi accretion does~\citep{Angles:2013}, and hence \simba\ can employ non-spherical feedback during the growth phase and yield black holes consistent with observed scaling relations.  (iii) At low-$\fedd$, Illustris-TNG randomises the direction of the jets at each timestep, rather than always ejecting jets perpendicular to the inner disk.  Our approach seems more physically motivated, since jets are not known to dramatically precess on timescales of $\sim$Myr (though they may occasionally do so over hundreds of Myr).  (iv) Illustris-TNG uses, at maximum, 200\% of the AGN bolometric luminosity (assuming a 10\% radiative efficiency), whereas for our model, the maximum is approximately a third of the bolometric luminosity.  Despite these and other minor differences, the use of two-mode AGN feedback as in Illustris-TNG and \simba\ seems to be a reasonably successful approach in state of the art AGN feedback models.

\subsubsection{X-ray feedback}

We include energy input into surrounding gas from X-rays off the accretion disk, as motivated and discussed in \citet{Choi:2012}.  Specifically, we compute the volume heating rate owing to X-rays following equation~12 of \citet{Choi:2012}, assuming (as they did) a radiative efficiency of 0.1.  We only apply this heating when jet mode is activated, as the lower velocity winds typically arise in more gaseous blue galaxies for which radiative losses would be severe~\citep{Best:2012}.  To be more explicit, we assume that more gas-rich galaxies are able to absorb and radiate away the X-ray energy, so we implement a gas fraction threshold such that we only apply X-ray heating if $f_{\rm gas} < 0.2$, where $f_{\rm gas}=M_{\rm gas}/M_*$ as computed by our galaxy finder, and we only include X-ray heating in galaxies with full-velocity jets. 

The X-ray heating is applied to gas within the black hole accretion kernel, scaled inversely with the square of the distance between the gas elements and the black hole, including Plummer softening based on the gas's smoothing length in order to mitigate large energy deposition in gas close to the black hole.  For non-ISM gas, we directly increase the gas's temperature according to the heating flux at the gas's position.   For ISM gas, because depositing such heat into a low-resolution, pressurised ISM as we assume in \simba\ would cool quickly and not be physically well motivated, we instead take half of the X-ray energy and apply it a radial outwards kick; the remainder is added as heat.  We further limit the total energy input in both kinetic and thermal forms to the overall available heating energy; if while looping over BH neighbors the X-ray energy input exceeds this value, then no further X-ray heating is done for that black hole at that timestep.  The X-ray heating has a fairly minimal effect on the galaxy mass function, but it provides an important additional energy input to more fully quench massive galaxies, as we discuss in \S\ref{sec:variants}.

\subsection{Black hole seeding and dynamics}\label{sec:seeds}

We use the on-the-fly FOF algorithm to seed black holes in galaxies dynamically during the simulation \citep[e.g.][]{DiMatteo:2008,Angles:2017}.  If a galaxy reaches a stellar mass $M_{*} > \gamma_{\rm BH} \times M_{\rm seed}$ and it does not already contain a black hole particle, then the star particle closest to the center of mass of the galaxy is converted into a black hole particle.  For our fiducial simulations, we employ $M_{\rm seed} = 10^4$\,M$_{\odot}/h$~and $\gamma_{\rm BH} = 3\times10^5$, which places black holes in galaxies with $M_{*} \gtrsim 10^{9.5}$\,M$_{\odot}$.  

This somewhat high stellar mass threshold for black hole seeding is motivated by recent simulations from the FIRE project, showing that stellar feedback strongly suppresses black hole growth in low mass galaxies by evacuating the nuclear gas reservoir on $<100$\,pc scales \citep{Angles:2017c}.  
%This was also found in EAGLE \citep{Bower:2017,McAlpine:2018} and {\sc Ramses}-based simulations~\citep{Dubois:2015,Habouzit:2017}.  
A qualitatively similar effect was also found in EAGLE~\citep{Bower:2017,McAlpine:2018} and {\sc Ramses}-based simulations~\citep{Dubois:2015,Habouzit:2017}, though their use of Bondi accretion may inhibit the growth of low mass black holes even in the absence of resolved stellar feedback (owing to the strong dependence $\dot{M}_{\rm Bondi} \propto M_{\rm BH}^{2}$).
Owing to poorer cosmological resolution as well as \simba's decoupled kinetic winds that explicitly avoids interaction of star formation feedback with ISM gas, \simba\ does not reproduce this effect self-consistently. Hence we simply seed black holes in the regime where they are expected to grow more efficiently.  We note that our results are insensitive to the exact choice of $M_{\rm seed}$ and stellar mass threshold~\citep{Angles:2015}.

%Since this effect is not resolved here\footnote{The EAGLE simulation reports effects of stellar feedback on black hole growth despite having similar resolution to \simba\ \citep{Bower:2017,McAlpine:2018}. For other simulations showing suppressed black hole growth in low mass galaxies, see e.g. \citet{Dubois:2015} and \citet{Habouzit:2017}.}, 

We assume that dynamical friction is efficient enough to maintain black holes near the host galaxy's center.  At every time step, black hole particles are repositioned to the location of the potential minimum within the FOF host group, if it is found within a distance $<4\times R_0$, where $R_{0}$ is the size of the black hole kernel used to compute the accretion rate.  The black hole particle velocity is then set to the center of mass velocity of the FOF group.  While current cosmological large volume simulations cannot self-consistently model the dynamics of black holes within galaxies, this algorithm is sufficient to capture the mass growth and feedback of ``well-behaved" central black holes \citep[see][for an attempt to include sub-grid dynamic friction for black holes in cosmological simulations]{Tremmel:2017}. Any two black holes located within $R_{0}$ are allowed to merge instantaneously if their relative velocity is lower than three times their mutual escape velocity.

\subsection{Dust production, growth and destruction}\label{sec:dust}

\simba\ includes a dust physics module to track the lifecycle of cosmic dust. In this implementation, dust is passively advected following the gas particles. This treatment is essentially accurate, as gas drag is usually able to decelerate grains on very short time scales especially when the radiative pressure is weak, so the drift cannot be resolved in our simulations. We additionally assume all dust grains have the same physical properties with a fixed radius $a\ =\ 0.1\ \mu m$. We ignore the active dust cooling, which will be applied in future work.

Dust is produced by condensation of metals from ejecta of SNe and AGB stars. We follow the prescription described in the work of \citet{Dwek:1998}, with updated condensation efficiencies based on recent studies. In the following, $m_{i,d}^j$ refers to the dust mass of the $i$th element (C, O, Mg, Si, S, Ca, Fe) produced by the $j$th stellar process (Type II SNe or AGB stars), whereas $m_{i,{\rm ej}}^j$ refers to the mass of ejecta from the $j$th process.

The mass of dust produced by AGB stars with a carbon-to-oxygen mass ratio C/O $>$ 1 is expressed as
\begin{equation}
m_{i,d}^{\rm AGB}=
\begin{cases}
\delta_{\rm C}^{\rm AGB} (m_{C,{\rm ej}}^{\rm AGB} - 0.75 m_{O,{\rm ej}}^{\rm AGB}), & i\ =\ {\rm C}\\
0, & {\rm otherwise,}
\end{cases}
\end{equation}
where $\delta_i^{\rm AGB}$ is the condensation efficiency of element $i$ for AGB stars. The mass of dust produced 
by AGB stars with a carbon-to-oxygen mass ratio C/O $<$ 1 is expressed as
\begin{equation}
m_{i,d}^{\rm AGB}=
\begin{cases}
0, &  i\ =\ {\rm C}\\
16 \sum \limits_{i=\rm{Mg,Si,S,Ca,Fe}} \delta_i^{\rm AGB} m_{i, {\rm ej}}^{\rm AGB}, & i\ =\ {\rm O}\\
\delta_i^{\rm AGB} m_{i, {\rm ej}}^{\rm AGB}, & {\rm otherwise,}
\end{cases}
\label{eq:2}
\end{equation}
where $\mu_i$ is the mass of element $i$ in atomic mass units. The mass of dust produced by Type II SNe is described as
\begin{equation}
m_{i,d}^{\rm SNII}=
\begin{cases}
16 \sum \limits_{i=\rm{Mg,Si,S,Ca,Fe}} \delta_i^{\rm SNII} m_{i, {\rm ej}}^{\rm SNII}, & i\ =\ {\rm O}\\
\delta_i^{\rm SNII} m_{i, {\rm ej}}^{\rm SNII}, & {\rm otherwise,}
\end{cases}
\label{eq:3}
\end{equation}
where $\sigma_i^{\rm SNII}$ is the condensation efficiency of element $i$ for Type II SNe.

We take a fixed dust condensation efficiency $\delta^{\rm AGB}_{i,\rm dust}=0.2$ based on the theoretical models of \cite{Ferrarotti:2006}. Guided by computations of \citet{Bianchi:2007}, we choose the dust condensation efficiency of Type II SNe $\delta^{\rm SN II}_{i,\rm dust}=0.15$ to match the low-metallicity end of the observed relation between dust-to-gas mass ratios (DGR) and gas-phase metallicities \citep{Remy-Ruyer:2014}. We omit the condensation of Type Ia SNe ejecta, as recent work suggests that Type Ia SNe are not significant sources of dust production \citep[see][]{Nozawa:2011,Dwek:2016,Gioannini:2017}. This is different from \citet{McKinnon:2016} and \citet{Popping:2017} where Type Ia SNe are assumed to have the same condensation efficiency as Type II SNe.

Once dust grains are produced, they can grow by accreting gas-phase metals. Derived by \cite{Dwek:1998}, the growth rate of grain radius can be expressed as:
\begin{equation}
\left( \frac{\diff M_{\rm dust}}{\diff t} \right)_{\rm grow}=\left( 1-\frac{M_{\rm dust}}{M_{\rm metal}} \right) {\left( \frac{M_{\rm dust}}{\tau_{\rm accr}} \right)},
\end{equation}
where $M_{\rm metal}$ is the total mass of dust and local gas-phase metals. Following \citet{Hirashita:2000} and \citet{Asano:2013} which assume the accretion is a two-body collisional process, the accretion time scale $\tau_{\rm accr}$ is 
\begin{equation}
\tau_{\rm accr} = \tau_{\rm ref} \left( \frac{\rho_{\rm ref}}{\rho_g} \right) \left( \frac{T_{\rm ref}}{T_g} \right) {\left(\frac{Z_\odot}{Z_g} \right)}.
\end{equation}
where $\rho_g$, $T_g$ and $Z_g$ are the local gas density, temperature and metallicity, respectively. 
$\rho_{\rm ref}$, $T_{\rm ref}$ and $Z_{\rm ref}$ are the reference values correspondingly. We take $\rho_{\rm ref} = 100$~H~atoms~cm$^{-3}$, $T_{\rm ref}=20$~K and $\tau_{\rm ref} = 10$~Myr. Inclusion of the multiplier $({Z_\odot}/{Z_g})$, unlike \citet{McKinnon:2017}, is integral to reproduce the observed relation between the dust to gas ratio and gas-phase metal abundance (\S\ref{sec:dustprop}).

Dust grains can be eroded by colliding with thermally excited gas especially in hot halos. A number of works have calculated the thermal sputtering rate in detail (e.g. \citealt{Barlow:1978,Draine:1979,Tielens:1994}). In 
this work, we adopt an analytic approximation of the growth rate of grain radii of \cite{Tsai:1995} (also adopted by \citealt{McKinnon:2017} and \citealt{Popping:2017}) described as
\begin{equation}
\left( \frac{\diff a}{\diff t} \right)_{\rm sp} = -\frac{a}{\tau_{\rm sp}},
\end{equation}
where the sputtering time scale
\begin{equation}
\begin{aligned}
\tau_{\rm sp} & = a \left| \frac{\diff a}{\diff t} \right|^{-1}\\ 
 &\sim (0.17{\rm Gyr})\left( \frac{a}{0.1 \mu m} \right) \left( \frac{10^{-27}{\rm g\ cm^{-3}}}{\rho_g} \right)\left[ \left( \frac{T_0}{T_g}\right)^{\omega}+1 \right],
\end{aligned}
\end{equation} 
where $\omega$~=~$2.5$ controls the low-temperature scaling of the sputtering rate and 
$T_0\ =\ 2 \times 10^6$~K is the temperature above which the sputtering rate flattens. The corresponding dust mass changes as
\begin{equation}
\left( \frac{\diff M_{\rm dust}}{\diff t} \right)_{\rm sp} = -\frac{M_{\rm dust}}{\tau_{\rm sp}/3}
\end{equation}

Because SN blast waves are not resolved in our simulations, we implement an additional dust destruction mechanism by SN shocks which enhance inertia and the thermal sputtering of dust grains \citep{Dwek:1980,Seab:1983, McKee:1987,McKee:1989}. We follow the prescription outlined by \citet{McKinnon:2016} in this work. The growth rate of the dust particle mass due to SN destruction is
\begin{equation}
\left( \frac{\diff M_{\rm dust}}{\diff t} \right)_{\rm de} = -\frac{M_{\rm dust}}{\tau_{\rm de}},
\end{equation}
where the characteristic time scale $\tau_{\rm de}$ is
\begin{equation}
\tau_{\rm de} = \frac{M_g}{\epsilon \gamma M_s},
\end{equation}
where $M_g$ is the local gas mass, $\epsilon = 0.3$ is the efficiency with which grains are destroyed in SN shocks \citep{McKee:1989}, $\gamma$ is the local SN II rate, and $M_s$ is the mass of local gas shocked to at least 100 km/s. Considering that our simulations are unable to resolve multi-phase ISM, we apply the Sedov-Taylor solution to a homogeneous medium of $n_{\rm H}=0.13$~H~atoms~cm$^{-3}$ (the minimum SF threshold density of our simulations) and obtain
\begin{equation}
M_s = 6800\ E_{\rm SNII,51} \left( \frac{v_s}{100\ {\rm km\ s^{-1}}} \right),
\end{equation}
where $E_{\rm SNII,51}$ is the energy released by a SN II in units of 10$^51$~erg, and $v_s\sim 100$~km~$s^{-1}$ is the shock wave speed.

We additionally destroy dust, as well as molecular hydrogen, completely in hot winds and during star formation (\S\ref{sec:code}) and {\color{black} in any gas that is impacted by AGN X-ray heating or jets (\S\ref{sec:bh}).  This is done instantaneously, with all dust mass and metals being returned to the gaseous phase.  Note that we do not do this for cold star-forming winds or AGN winds in the high-Eddington mode, so these outflows carry molecular gas and dust out of the galaxy.  We leave for future work an investigation into whether this reproduces observations of AGN-driven molecular outflows~\citep[e.g.][]{Sturm:2011} and circum-galactic dust~\citep[e.g.][]{Peek:2015}.}

\subsection{Runs and analysis}

The primary \simba\ runs have $1024^3$ dark matter particles and $1024^3$ gas elements.  We are running four volumes: $100\hmpc$ down to $z=0$, $50\hmpc$ to $z=1$, $25\hmpc$ to $z=2$, and $12.5\hmpc$ to $z=5$.  All runs have identical input physics, begin at $z=249$, and assume a \citet{Planck:2016} concordant cosmology of $\Omega_m=0.3$, $\Omega_\Lambda=0.7$, $\Omega_b=0.048$, $H_0=68\kmsmpc$, $\sigma_8=0.82$, and $n_s=0.97$.  Other parameters such as the minimum gravitational softening length and mass resolutions are listed in Table~~\ref{tab:sims}.  In this paper we will only present results from the main $100\hmpc$ run, as the other runs are at various stages of completion.

We will also explore parameter space and compare to our previous \mufasa\ simulations using $50\hmpc$ runs with $2\times 512^3$ particles that match \mufasa's size.  We run a full physics \simba\ simulation at this resolution, and in order to directly assess the impact of our new quenching feedback modules, namely jet and X-ray feedback, we also run a ``No-jet" simulation where these modules are turned off, and a "No-Xray" run where jets are kept on but X-ray feedback is turned off.  All other input physics in these runs, including stellar feedback and radiative mode black hole feedback, remains identical to that in \simba.

\begin{figure*}
	\includegraphics[width=\columnwidth]{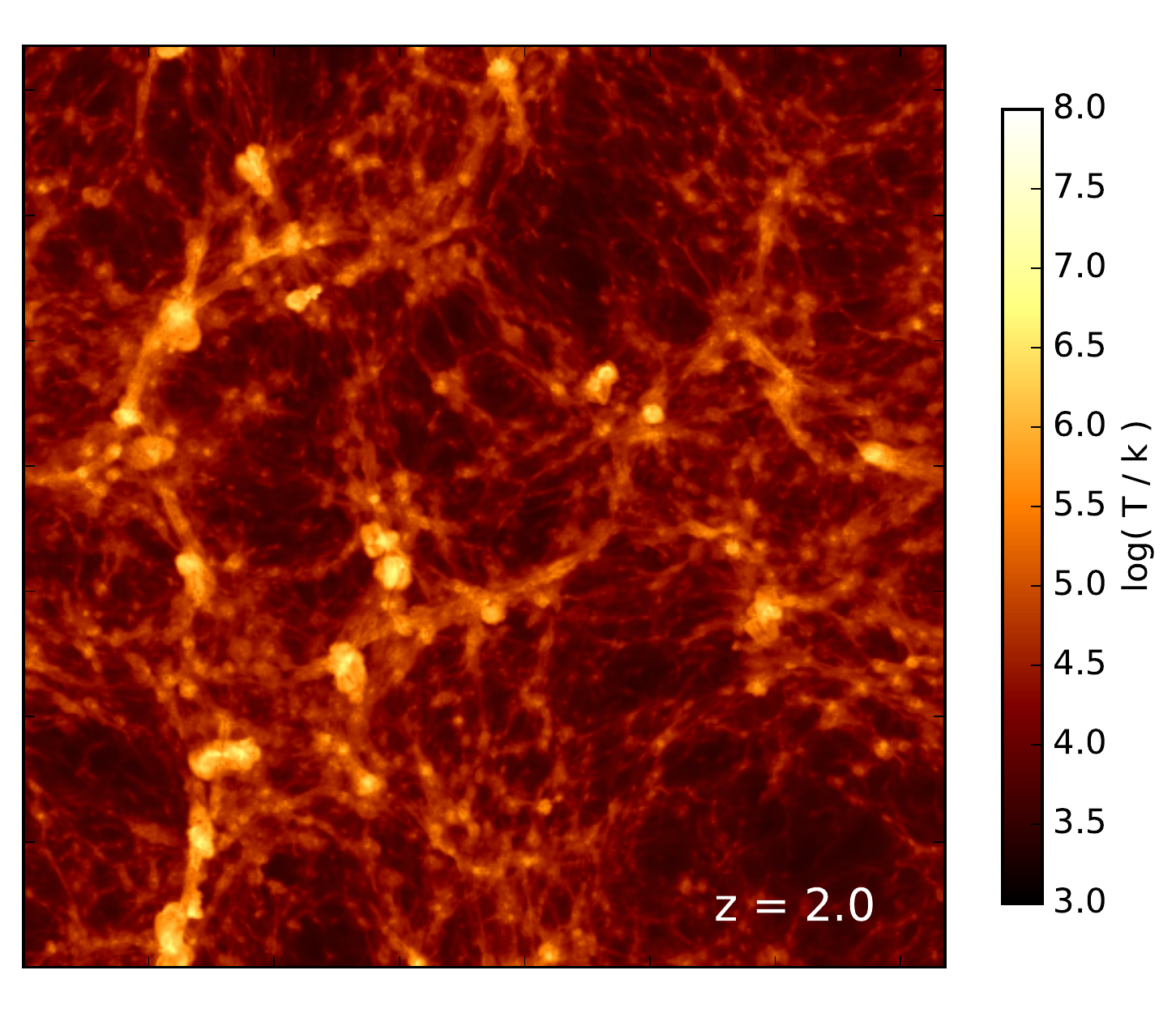}
    \includegraphics[width=\columnwidth]{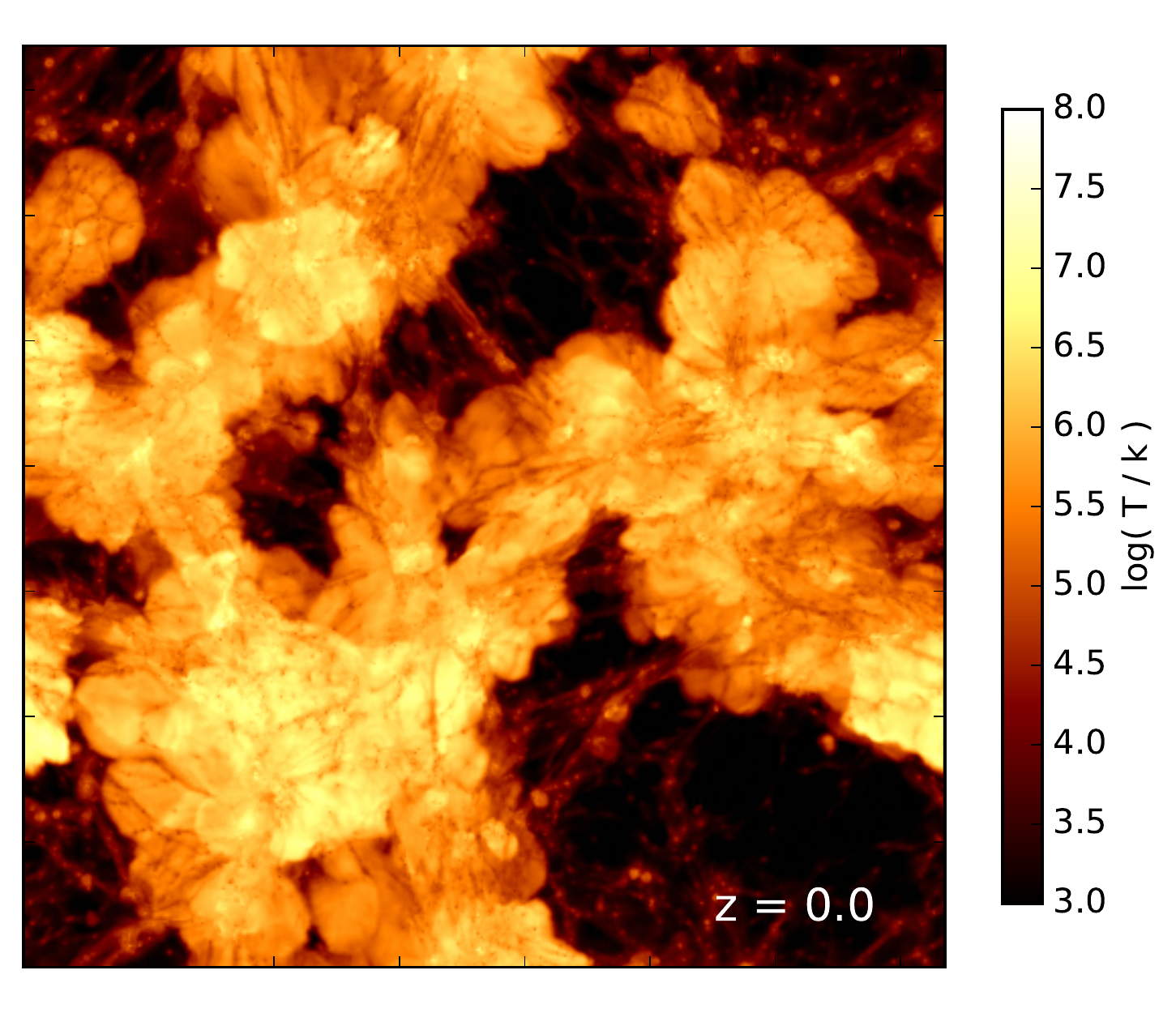}
    \caption{Temperature map projected through a random 10~Mpc/h slice from a 50 Mpc/h \simba\ volume, at $z=2$ (left) and $z=0$ (right).  At $z=2$, warm-hot gas traces large-scale filaments, with energetic bipolar outflows owing to jets evident from the nodes where the most massive galaxies and black holes reside.  At $z=0$, high-speed AGN outflows have shocked the IGM gas throughout much of this volume to well beyond the virial radii of halos, with cooler dense filamentary structures penetrating the hot gas.}
    \label{fig:tmap}
\end{figure*}

\begin{table*}
	\centering
	\caption{The \simba\ simulation suite.}
	\label{tab:sims}
	\begin{tabular}{lcccccc} % four columns, alignment for each
		\hline
		Name & $L_{\rm box}^a$ & $\epsilon_{\rm min}^b$ & $z_{\rm end}^c$ & $m^d_{\rm gas}$ & $m^e_{\rm DM}$ & $M^f_{\rm *,min}$\\
		\hline
		m100n1024 & 100 & 0.5 & 0 & $1.82\times 10^7$ & $9.6\times 10^7$ & $5.8\times 10^8$\\
		m50n1024 & 50 & 0.25 & 1 & $2.28\times 10^6$ & $1.2\times 10^7$ & $7.3\times 10^7$\\
		m25n1024 & 25 & 0.125 & 2 & $2.85\times 10^5$ & $1.5\times 10^6$ & $9.1\times 10^6$\\
        m12.5n1024 & 12.5 & 0.0625 & 5 & $3.56\times 10^4$ & $1.88\times 10^5$ & $1.14\times 10^6$\\
		\hline
	\end{tabular}
    \\$^a$ Box length in comoving $\hmpc$.
    \\$^b$ Minimum gravitational softening length in comoving $\hkpc$.
    \\$^c$ Ending redshift (all begin at $z=249$).
    \\$^d$ Initial gas element mass resolution in $M_\odot$.
    \\$^e$ Dark matter particle mass resolution in $M_\odot$.
    \\$^f$ Minimum stellar mass of a resolved galaxy in $M_\odot$.
\end{table*}

To analyse the simulation outputs, we employ a suite of tools as described below.  First, 
galaxies are identified using a friends-of-friends galaxy finder, assuming a spatial linking length of 0.0056 times the mean inter-particle spacing (equivalent to twice the minimum softening length).  In our tests, this gives very similar results to the more comprehensive Spline Kernel Interpolative Denmax (SKID) galaxy finder.  Galaxy finding is applied to all stars and black holes plus all gas elements with a density above the minimum SF threshold density of $n_H>$0.13~H~atoms~cm$^{-3}$; this captures all the stars and molecular gas in galaxies.  Black holes are assigned to the galaxy to which they are most gravitationally bound; large galaxies can have many black holes.  We take the central black hole to be the most massive black hole in the galaxy, and use this when we discuss black hole masses.  In most cases, the other black holes are very small and add no significant black hole mass compared to the central one.

Because significant amounts of neutral hydrogen can lie in an extended configuration beyond the star-forming region of galaxies, we assign \ion{H}{I} to galaxies in a separate step.  To do this, we consider all gas elements with \ion{H}{I} fractions above 0.001, and assign them to the galaxy to which they are most gravitationally bound, i.e. its kinetic energy relative to the galaxy's center of mass velocity minus the potential energy from the galaxy at the gas element's location is minimised.

Halos are identified on the fly during the simulation run using a 3-D friends-of-friends algorithm within {\sc Gizmo}, which is identical to the one in \gad\ written by V. Springel.  The linking length is taken to be 0.2 times the mean inter-particle spacing.  We do not identify or consider sub-halos in this work.

Galaxies and halos are cross-matched in post-processing using the {\sc yt}-based package {\sc Caesar}, which outputs a single hdf5 catalogue containing all galaxy and halo information with many key properties pre-computed, as well as particle lists of individual galaxies and halos so that any other properties can be computed via user-written {\sc python} scripts.

%Galaxy magnitudes are computed using {\sc PyLoser}\footnote{\tt https://pyloser.readthedocs.io/en/latest}, which is a python version of {\sc Loser} (Line Of Sight Extinction by Ray-tracing) described in \citet{Dave:2017b}.  {\sc PyLoser} employs the Flexible Stellar Population Synthesis~\citep[FSPS;][]{Conroy:2009,Conroy:2010} model to compute the spectrum of each star particle, and then extincts that spectrum based on the line of sight gas metal column density in the chosen direction.  The summed spectra of all star particles in the galaxy represents the galaxy's spectrum, which is redshifted and convolved with bandpasses as desired.  An extinction law must be chosen, for which we use a Milky Way law~\citep{Fitzpatrick:2007} for galaxies with specific star formation rates (sSFR) below 0.1~Gyr$^{-1}$, and a \citet{Calzetti:2000} law for higher sSFR galaxies.  We include nebular emission lines as modeled in FSPS, but this has no significant effect for the results presented here.   {\sc PyLoser} also has the ability to make galaxy spectra, images, and data cubes, both for photometry as well as gas emission line properties, but we do not employ these features in this work.

\begin{figure}
	\includegraphics[width=0.45\textwidth]{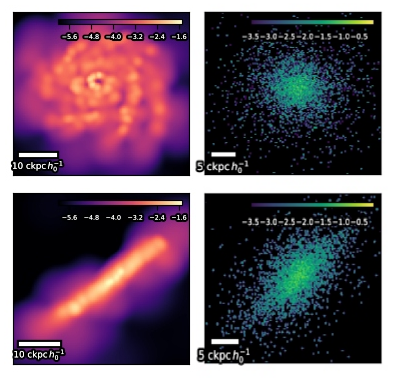}
	\includegraphics[width=0.45\textwidth]{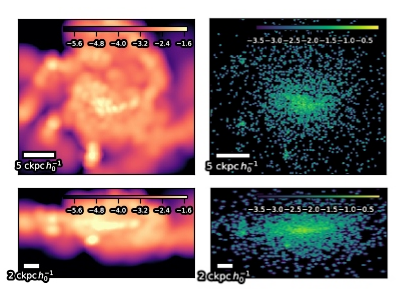}
	\vskip-0.1in
    \caption{Examples of the molecular gas (left) and stellar (right) surface density distributions in star-forming disk galaxies with $M_*\approx 4.7\times 10^{10}M_\odot$ at $z=0$ (top four panels) and $z=2$ (bottom four panels), showing face-on and edge-on views.  At $z=0$ there is a thin, well-ordered disk in both \HI\ and H$_2$, while the high-$z$ galaxy is clumpier and thicker.}
    \label{fig:disks}
\end{figure}

All results shown here are obtained from the {\sc Caesar} catalogs generated from simulation snapshots at specified redshifts.  We output 151 snapshots to $z=0$, 105 to $z=1$, and 78 to $z=2$.  Each snapshot is $\approx$250~GB in size, and the {\sc Caesar} catalogues are typically $\sim$15~GB each.

\section{Results}\label{sec:results}

\begin{figure*}
	\includegraphics[width=6.2in]{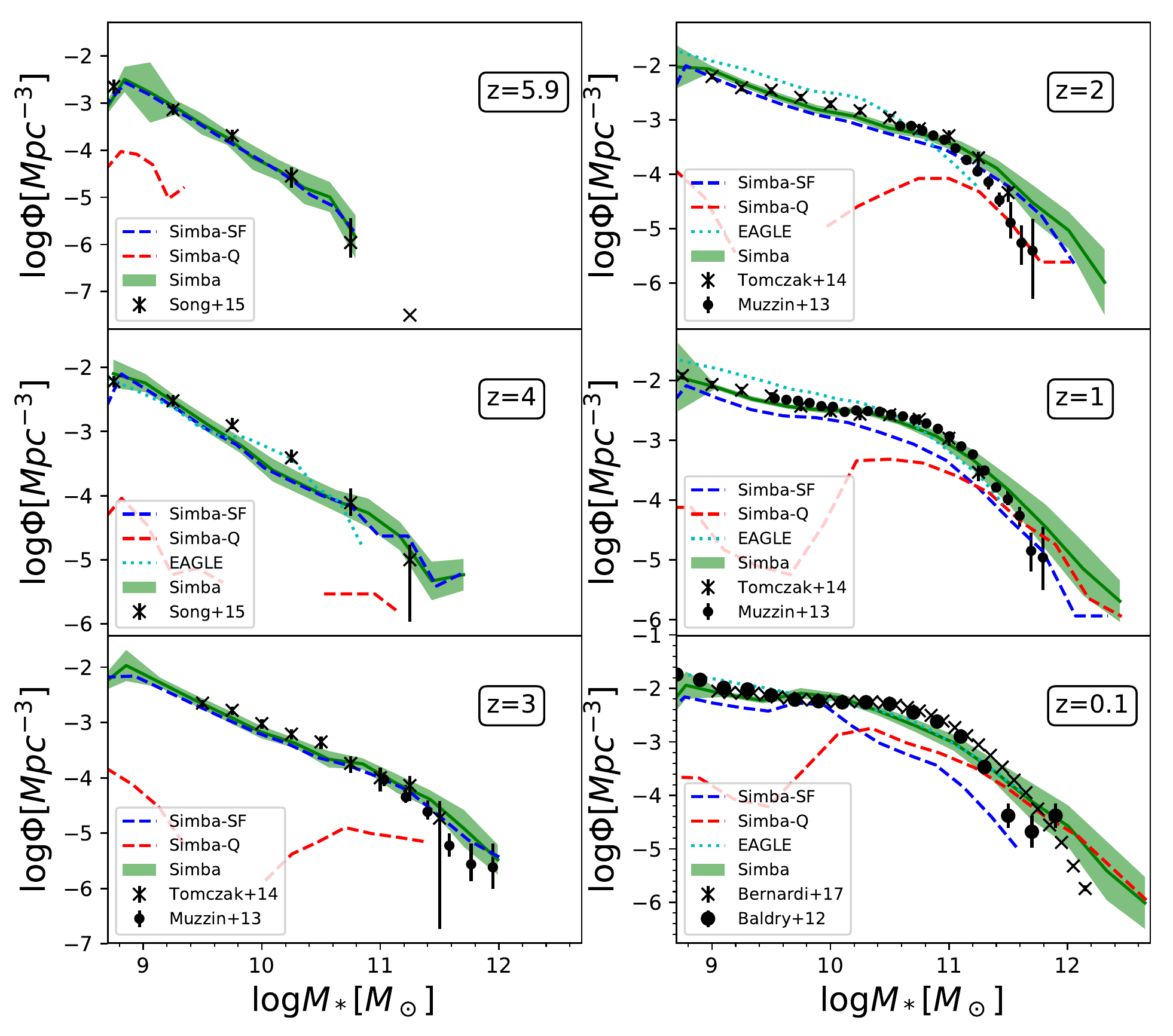}
	\vskip-0.1in
    \caption{Stellar mass function evolution from $z=6\rightarrow 0$, compared to observations as indicated in the legends.  Green band shows the results from all \simba\ galaxies, with the spread computed from jackknife resampling the 8 simulations sub-octants. Red and blue dashed lines show the mass functions of central galaxies below and above sSFR$=10^{-1.8+0.3z}$Gyr$^{-1}$, respectively.  Cyan dotted line shows the results from EAGLE for comparison.}
    \label{fig:mfwind}
\end{figure*}

In this section we provide a comprehensive suite of predictions for \simba\ for a range of key global galaxy properties.  The purpose is to ascertain how well \simba\ reproduces observed galaxy stellar and gas properties that have historically provided stringent constraints on feedback models in previous simulations, and thereby demonstrate the suitability of \simba\ as a platform to study detailed galaxy evolution.

To begin, we show in Figure~\ref{fig:tmap} a projected temperature map from the (50 Mpc/h)$^3$, $512^3$ \simba\ simulation. The slice shown is 10~Mpc/h thick, and is arbitrarily chosen to contain representative structures in the volume.  At $z=2$, the familiar Cosmic Web is evident as traced out by warmer gas arising from mild shock heating on filamentary structures.  Closer inspection reveals the earliest AGN jets becoming active, with characteristic bipolar outflows that are typically perpendicular to the large-scale filaments.  But these large early black holes are sparse, and most of the IGM is unaffected by feedback.  In contrast, by $z=0$ (right panel), a significant volume of the IGM has been heated to high temperatures from AGN feedback, and the hot bubbles encroach upon regions untouched by AGN feedback containing the canonical warm filaments.  These bubbles are reasonably spherical since they arise from clustered massive galaxies, each one ejecting jets that are relatively stable in direction but overlap quickly with neighboring outflows.  In some cases individual bipolar jets and the resulting bow shocks can still be picked out. Such a dramatic impact on the IGM may have significant consequences for the ionisation state of diffuse neutral hydrogen~and the statistics of Lyman alpha forest absorbers \citep{Kollmeier:2014}, as well as the diffuse IGM pressure measurable via the Sunyaev-Zel'dovich effect~\citep{Lim:2018}; we will explore these in future work.  In this paper, we focus on the demographics of the galaxy population predicted by \simba.

Figure~\ref{fig:disks} shows some examples of individual galaxies. We choose a Milky Way-sized disk galaxy at $z=0$, with $M_*\approx 4.7\times 10^{10}M_\odot$ and SFR$=1.3\ M_\odot$yr$^{-1}$, and show the face-on (upper row) and edge-on (lower row) views, in both H$_2$ surface density (left) and stellar mass surface density (right).  The $z=2$ galaxy shown in the bottom four panels has essentially the same $M_*$, but with SFR$=45\ M_\odot$yr$^{-1}$ that is typical of a main sequence galaxy at Cosmic Noon. The $z=0$ disk is a grand design spiral, with a thin cold gas distribution. There is a small central hole in cold gas that owes to the AGN feedback from its $6\times 10^7M_\odot$ black hole accreting at $0.005 M_\odot$yr$^{-1}$.  The stellar distribution does not show the spiral structure owing to the relatively low resolution of \simba, compared to zooms or higher-resolution simulations such as Illustris-TNG and EAGLE.
The $z=2$ system shows more prominent star forming clumps and a thicker gas distribution, and is overall more compact (note the scale bar). While \simba's numerical resolution smooths out many of the detailed internal features, this shows that it still produces galaxies that have features broadly like star-forming disk galaxies in the real Universe.  We do not show more massive quenched examples, but as expected they tend to be elliptical in their stellar morphology, with little cold gas.

\subsection{Galaxy stellar mass functions}

Since galaxies are a collection of stars, the most basic property of a galaxy is its stellar mass.  Given that the concordance cosmological model strongly constraints the halo mass function, the galaxy stellar mass function (GSMF) thus characterises the efficiency by which halos convert their baryons into stars.  It is well established that (under the abundance-matching ansatz) the stellar-to-halo mass ratio drops quickly to low and high masses away from the peak at $L^*$~\citep[e.g.][]{Moster:2013,Behroozi:2013}, and current models attribute this to self-regulation by star formation-driven feedback below $L^*$ and quenching of galaxies due to AGN feedback above $L^*$~\citep{Somerville:2015}.  Since the GSMF is reasonably well measured over much of cosmic time~\citep[albeit with non-trivial systematic uncertainties;][]{Mobasher:2015}, it represents a stringent test for the key feedback modules of a galaxy formation model.  Indeed, simulations these days including \simba\ tend to use the $z=0$ GSMF as a primary constraint to tune feedback models.

Figure~\ref{fig:mfwind} shows the GSMF at $z=0.1,1,2,3,4,6$ from \simba\ (green lines). Observational data is shown at $z=0$ from \citet{Bernardi:2017}.  At $z=1,2,3$ we show observations from \citet{Tomczak:2014} combining CANDELS and zFOURGE data, while at $z=4,6$ we show observations based on CANDELS from \citet{Song:2016}.  We also show the GSMF of central galaxies only, subdivided into star-forming (SF) and quenched (Q) samples at a specific SFR$=10^{-1.8+0.3z}$Gyr$^{-1}$.  Error bars are shown from jacknife re-sampling over eight simulation sub-octants. Finally, we show the results from the EAGLE simulation as the dotted cyan line at selected redshifts.

\simba\ produces generally good agreement with the observed GSMFs at all redshifts, overall comparably well to EAGLE. There is excellent agreement at $z\geq 3$, especially given the systematic uncertainties in stellar mass determinations at higher redshifts \citep{Mobasher:2015}.  At $z=2$, there starts to be a slight excess at the massive end in \simba. This may owe to insufficient quenching of the most massive galaxies, or may represent an underestimate of the observed GSMF owing to selection effects in the rest-optical surveys used for the GSMF determinations which can miss massive dusty galaxies.

At lower redshifts, there is a clear truncation at the massive end, but a mild overproduction of the most massive galaxies remains all the way to $z=0$.  Like EAGLE, 
\simba\ {\color{black} under-predicts the GSMF around $M^\star$ by a factor of up to two}; this was an advantage of \mufasa\ that is unfortunately not retained in \simba.  This highlights that it continues to be a challenge to achieve such a sharp turndown in the GSMF using a physically-motivated AGN feedback model.

The overproduction of the most massive galaxies could owe to a number of effects. First off, there are numerical uncertainties in quantifying the most massive systems, because they tend to have large extended envelopes of stars and many satellites that, owing to poor resolution, can be overmerged into the central object either during the dynamical evolution or during the post-processing galaxy finding stage.  These tend to artificially boost the mass in the simulated massive galaxies. One way to mitigate this is to compare the stellar mass within fixed apertures to data, which \citet{Schaye:2015} showed using EAGLE can significantly reduce the mass of $M_*\ga 10^{11}M_\odot$ objects. There are also increased observational uncertainties at the massive end.  For instance, it is a matter of debate as to how much of the surrounding stars should be classified as part of the central galaxy and how much should be intracluster light; this can strongly impact the stellar mass~\citep{Kravtsov:2018}. There is also the issue of the stellar initial mass function (IMF) -- stellar population~\citep{Conroy:2013} and dynamical~\citep{Cappellari:2013} studies suggest that the most massive galaxies have bottom-heavy IMFs relative to Milky Way-like galaxies, which can result in the stellar mass being underestimated by a factor of 2 or more for the most massive systems. Finally, there is an issue particular to this \simba\ run -- it turns out, in the 100$\hmpc$ volume, by $z=0$, the largest halo has a virial mass of $M_{\rm halo}=1.16\times 10^{15}M_\odot$, which is larger than expected by about 50\% for its volume; this may contribute to the excess of the very most massive galaxies. Hence although at face value there is some disagreement at the massive end in comparing \simba\ with recent observations, more work must be done to determine whether these discrepancies reflect a significant failing of \simba's input physics.

\begin{figure*}
    \subfloat{\includegraphics[width=0.45\textwidth]{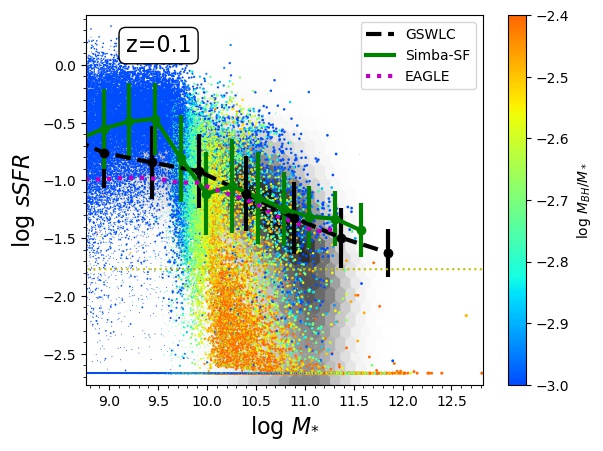}}
    \subfloat{\includegraphics[width=0.45\textwidth]{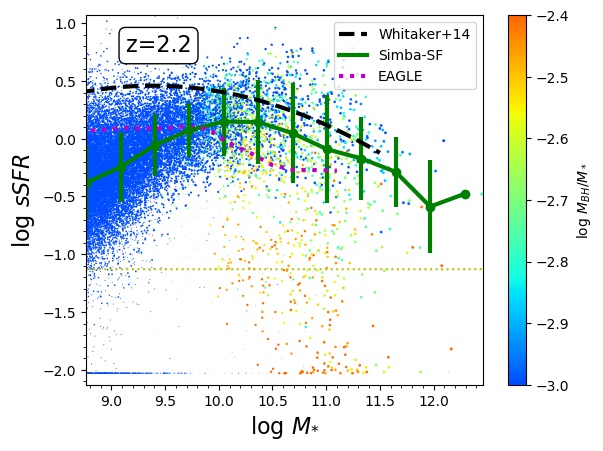}}
    \caption{Star formation rate--stellar mass relation at $z\approx0$ (left) and $z\approx 2$ (right).  Points show \simba\ galaxies, colour-coded by their black hole to stellar mass ratio.  The thick green line shows the running median to star-forming galaxies (i.e. above the horizontal dotted yellow line). Observations at $z=0$ from GSWLC-X2 are shown as the grey hexbins, with the black dashed line showing the median to galaxies using the same sSFR cut as shown for \simba.  The errorbars show the $1\sigma$ spread around the running median value, typically $0.3-0.4$~dex.  At high-$z$ we show the black dashed line as the best-fit relation for $2<z<2.5$ galaxies from \citet{Whitaker:2014}. Results from EAGLE are shown as the magneta dotted line for comparison.  \simba\ reproduces the star-forming main sequence at both redshift reasonably well, especially accounting for systematics in high-$z$ sSFR determinations, though in small galaxies it appears to overpredict the SFR at $z\sim 0$ and underpredict at $z\sim 2$}
    \label{fig:mainseq}
\end{figure*}

Examining the SF vs. Q samples, we see that massive quenched galaxies begin to appear in significant numbers at $z\ga 2$.  By $z=1$ they outnumber the SF galaxies among the most massive galaxies with $M_*\ga 10^{11}M_\odot$, and by $z=0$, they dominate at $M_*\ga 2\times 10^{10}M_\odot$. The quenched population grows quickly at low redshifts, and the number of massive star-forming galaxies drops quickly since $z\sim 1$, in broad agreement with observations~\citep[e.g.][]{Bell:2004}. There are a few very small quenched centrals, but this is likely an artifact of the friends-of-friends halo finder.  

In summary, to within systematic uncertainties, \simba\ produces a GSMF that is in quite good agreement with observations across most of cosmic time, with the possible exeption of the $z=0$ massive end. \simba\ passes this primary check at a level comparable to that seen for  \mufasa, EAGLE, and 
Illustris-TNG~\citep{Pillepich:2018}.  In no small part, this owes to these various models tuning feedback parameters to match such data, but even the fact that such a tuning is now possible is a recent and important step forward for cosmological hydrodynamic simulations. It does mean that the growth of galaxies' stellar component over time is no longer a strong discriminant between current galaxy formation models.  Instead, for this we must rely on the many other predicted observables that are not used to tune the models.  We now examine some of these predictions for \simba.

\subsection{Star formation rate--stellar mass relation}

Another key barometer of galaxy formation models is the star formation rate--stellar mass (SFR$-M_*$) relation.  Unlike the GSMF that is often used as a primary constraint on models, SFR$-M_*$ is not, making it more of a true prediction of models.  The SFR$-M_*$ relation consists of a star-forming ``main sequence" of galaxies, and a population of quenched galaxies falling below the main sequence that dominates at high masses at later epochs.  Getting the balance of these populations in accord with observations over cosmic time, as well as predicting their growth rates, has traditionally been difficult to reproduce in cosmological simulations.

During Cosmic Noon, it has long been seen that cosmological models tend to underpredict the main sequence amplitude~\citep{Daddi:2007,Dave:2008,narayanan12a,Sparre:2015,Somerville:2015}, typically by a factor of $2-3$.  Fixing this requires rather substantially changing the star formation histories, not just the overall SFRs, since a multiplicative constant on the SFR will tend to move galaxies along the relation rather than increase its amplitude.  There are also potential observationally-oriented systematics that may be overestimating the SFR owing to one or more of many possible factors, such as galaxies being dominated by harder-ionising stellar populations at high-$z$, or having a more top-heavy initial mass function.

Figure~\ref{fig:mainseq} shows the specific SFR$-M_*$ relation at $z=0.1$ (left) and $z=2.3$ (right) for \simba\ galaxies.  {\color{black} The SFRs are computed as instantaneous SFRs from the gas elements, which corresponds well to the SFR computed from young star particles when averaged over several tens of Myr.} The running median (green curve) includes star-forming galaxies only (\simba-SF), defined as before by sSFR$>10^{-1.8+0.3z}$Gyr$^{-1}$ (dotted horizontal yellow line).  {\color{black} The error bars show the $1\sigma$ spread around the median value in each bin.} Points are colour-coded by the ratio of black hole to stellar mass, with magenta points having higher $M_{BH}/M_*$; points at $M_*\la 10^{10}M_\odot$ in cyan have no or very small growing black holes, as we will discuss later.  Galaxies with very low or zero SFR are plotted near the bottom for visibility. Observations at low-$z$ are shown from the GALEX-SDSS-WISE Legacy Catalog\citet[GSWLC;][]{Salim:2016,Salim:2018}, shown as grey hexbins, and the running median to the star-forming galaxies with the same criterion as above is shown as the black dashed line{\color{black}, along with error bars showing the $1\sigma$ spread around the median}. At high-$z$, we show the median sSFR$-M_*$ relation measured for $2<z<2.5$ galaxies by \citet{Whitaker:2014}. Finally, results from EAGLE are shown as the magenta dotted line.

At $z=0$ (left panel), \simba\ nicely reproduces the observed GSWLC main sequence slope and amplitude at $M_*\ga 10^{10}M_{\odot}$.  Below this mass, \simba\ shows noticeably higher SFRs.  
This mass corresponds to the onset of massive black holes, as shown by the growing number of magenta-coloured points with higher black hole mass for their $M_*$.  Indeed, there is a very strong trend that the galaxies that are quenching are specifically the ones with a high $M_{BH}/M_*$ ratio; massive galaxies left on the main sequence at $z\sim 0$ in \simba\ are only those that for some reason have not grown their black hole as rapidly.  {\color{black}A similar trend is seen in EAGLE~\citep{Matthee:2019}, which arises owing to a spread in halo formation times~\citep{Davies:2019}.}  We will investigate the detailed reasons for this dichotomy in \simba\ in future work, but for now we note the tight connection between quenching and black holes already appearing in \simba, which will be a recurring theme throughout this paper.  {\color{black} The average slope of sSFR$-M_*$ for star-forming galaxies over the entire mass range plotted is $-0.27$, which is in reasonable agreement with observations~\citep[e.g.][]{Noeske:2007,Speagle:2014}.  The scatter around the main sequence in \simba\ is $0.3-0.4$~dex, with a mild tendency to drop with $M_*$; this is very comparable to that seen in the GSWLC data.} 

\begin{figure}
	\includegraphics[width=0.48\textwidth]{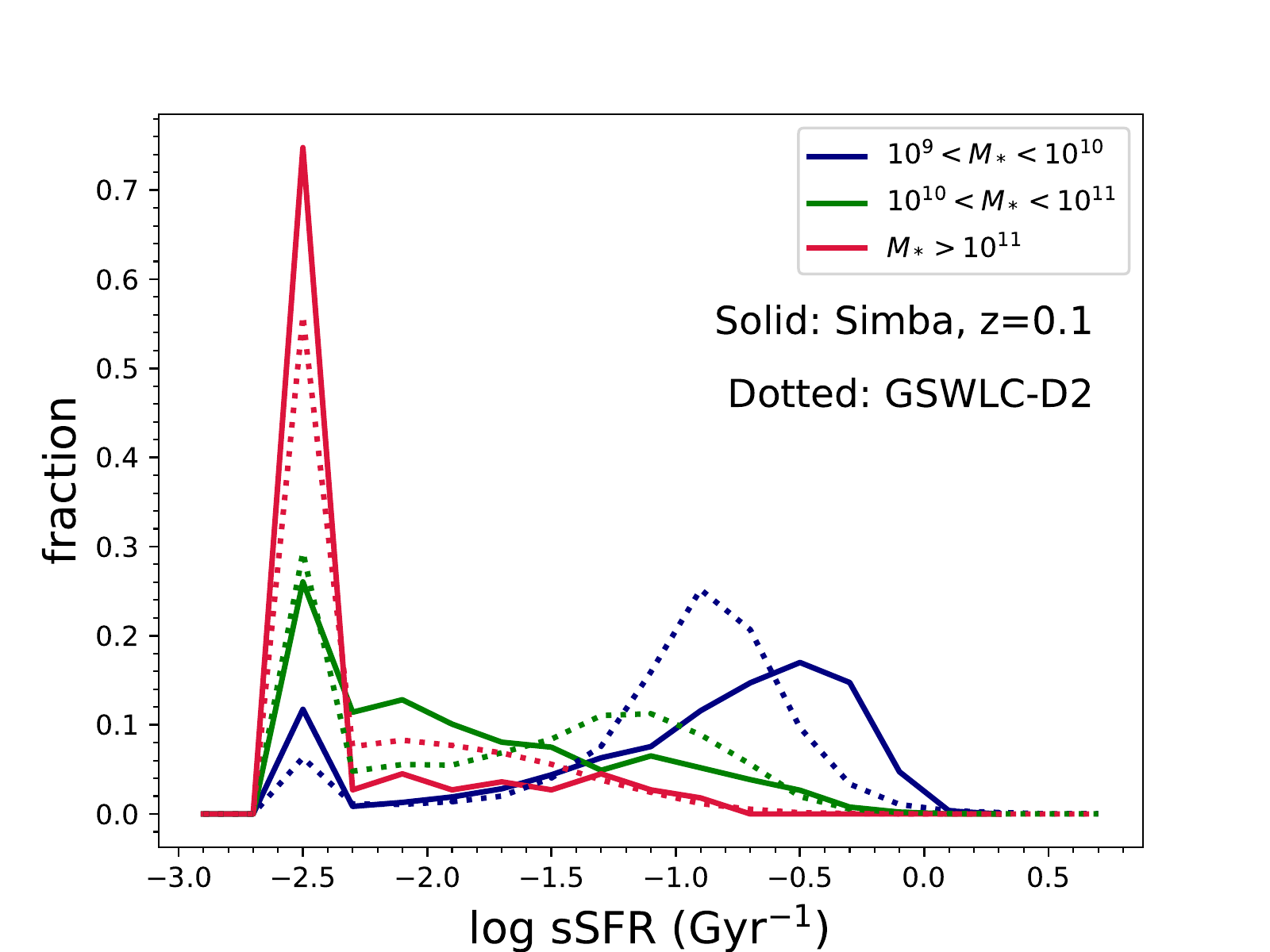}
	\vskip-0.05in
    \caption{Histogram of sSFR in three bins of stellar mass.  Solid lines show the results for \simba\ at $z=0.1$, while dotted lines show $z\sim0.1$ observations from GSWLC-D2.  All galaxies with sSFR$<10^{-2.5}$~Gyr$^{-1}$ are placed in the lowestmost bin.
    There is good agreement, particularly in the quenched fractions in more massive galaxies, though \simba\ produces somewhat too high sSFRs at low-$M_*$.}
    \label{fig:mainseqhist}
\end{figure}

At $z=2.2$ (right panel), the \simba\ main sequence generally tracks the observed one from \citet{Whitaker:2014}, but is low in amplitude by $\approx\times 2$.  This continues the trend in models that the main sequence at Cosmic Noon remains too low, though not quite as strongly as in some previous models. However, \citet{Leja:2018} points out that more sophisticated SED fitting applied to the latest datasets can lead to a systematic increase in the inferred $M_*$ while lowering the SFR that results in a combined $\approx 0.3$~dex lower sSFR compared to previous determinations. If confirmed, then at face value this would bring \simba's (and other models') simulated main sequence into agreement with $z\sim 2$ observations at long last.

{\color{black} Finally, we show in Figure~\ref{fig:mainseqhist} histograms of the specific SFR, broken up into mass bins of $10^9<M_*<10^{10}M_\odot$, $10^{10}<M_*<10^{11}M_\odot$, and $M_*>10^{11}M_\odot$.  Solid lines show the results from \simba\ at $z=0.1$, while dashed lines show identically selected galaxies from the GSWLC-D2 catalog~\citep{Salim:2018}.  Galaxies with sSFR$\leq 10^{-2.5}$~Gyr$^{-1}$ have been placed in the lowest sSFR bin.  There is an overall bimodal distribution, with low-mass galaxies being predominantly star-forming, while massive galaxies are almost uniformly quenched.  There is impressive agreement in the lowest sSFR bin, showing that \simba\ well reproduces the quenched fractions at various $M_*$.  However, the low-mass galaxies in \simba\ have somewhat too high sSFR values, reflecting the same excess as seen in Figure~\ref{fig:mainseq}.
}

In summary, \simba\ generally reproduces the main sequence of star-forming galaxies as seen at $z\approx 0,2$, to within current systematic uncertainties. Potential disagreements from data lie mostly at lower masses, where the observations are less certain and more subject to selection effects.  The success at $M*\ga 10^{10}M_\odot$ is encouraging because it suggests that the balance of quenching and quenched galaxies in this transition mass near $M^\star$ is being reproduced roughly correctly in \simba.  There is also a strong trend that quenched galaxies at a given $M_*$ tend to have larger fractional black hole masses, which is an interesting prediction that can be tested in future samples of black hole mass measurements in sub-$M^\star$ galaxies.

{\color{black}
\subsection{Global star formation rate evolution}

The evolution of the cosmic SFR density (SFRD) has long been a key test for cosmological galaxy formation models.  While proper model comparisons to data can be challenging owing to the variety of different selection effects used to measure SFR over time, the recent compilation by \citet{Madau:2014} has provided a homogenised database for the SFRD that can be more robustly compared.

\begin{figure}
	\includegraphics[width=0.48\textwidth]{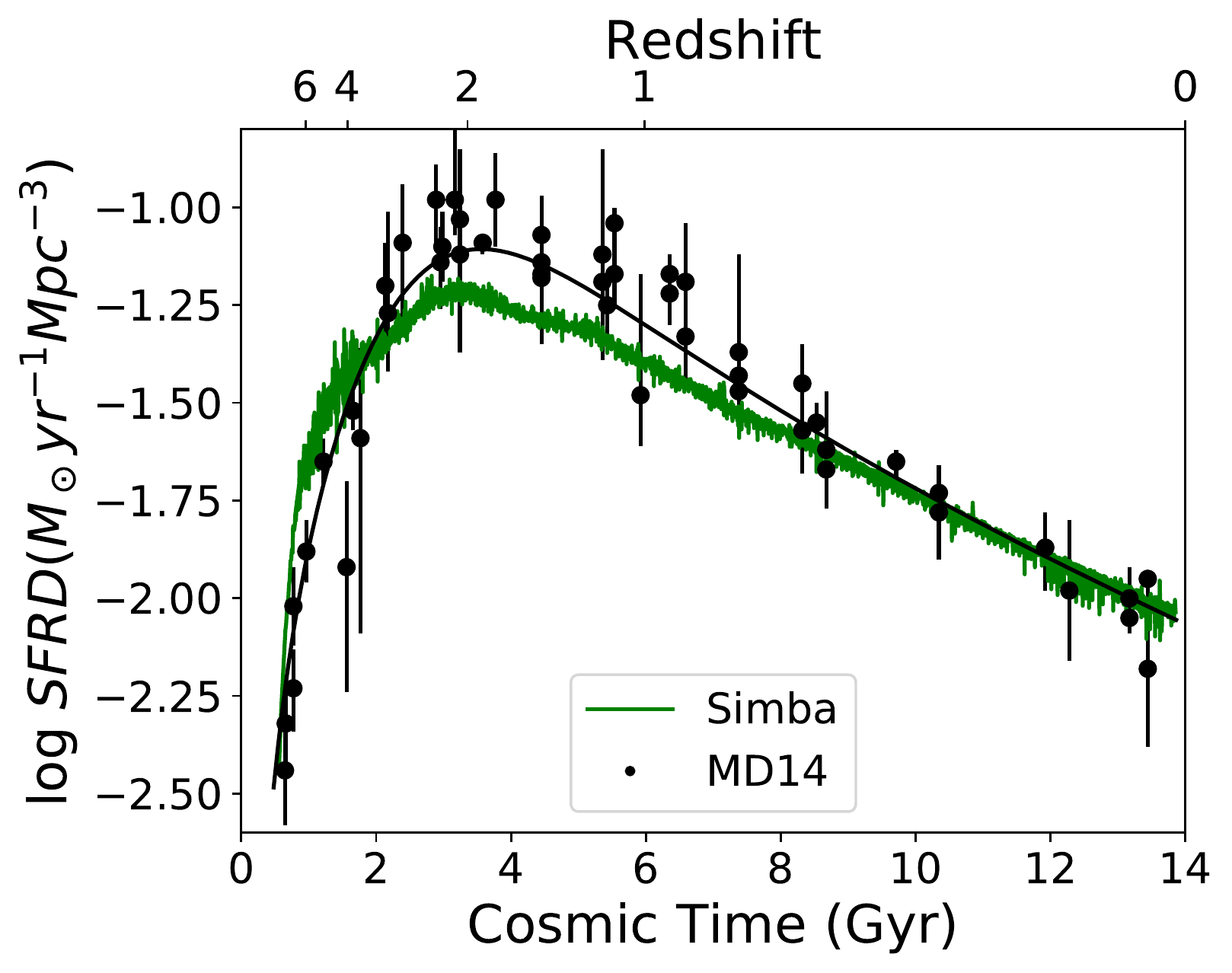}
	\vskip-0.1in
    \caption{Star formation rate density evolution versus age of the Universe in \simba\ (curve), compared to the observational compilation of \citet{Madau:2014} (black points and best-fit line).}
    \label{fig:madau}
\end{figure}

Figure~\ref{fig:madau} shows the comparison of the cosmic SFRD as a function of cosmic age in \simba\ versus the \citet{Madau:2014} compilation.  The \simba\ SFRD values include all the star formation in the volume at each epoch, but we have checked that including only star formation in resolved galaxies ($M_*>5.8\times 10^8 M_\odot$) makes a negligible difference.

The overall shape of the predicted SFRD versus time is in good agreement with observations.  \simba\ matches the preset-day SFRD very well, and generally reproduces the order-of-magnitude rise in SFRD towards the peak at $z\sim 2$. There is a slight tendency for \simba\ to form more stars globally at earlier epochs, with the peak shifted very slightly towards higher redshift compared to the best-fit line from \citet{Madau:2014}.  The peak SFRD at $z\sim2$ is also slightly lower than observed, following the trend shown in Figure~\ref{fig:mainseq} that \simba\ has slightly lower main sequence than observed.  Despite these minor differences, the overall shape and amplitude is in very good agreement with observations, comparable to the agreement seen versus other recent simulations such as Illustris-TNG~\citep{Pillepich:2018}.

}

\subsection{Neutral and molecular gas fractions}

\begin{figure}
	\includegraphics[width=3.5in]{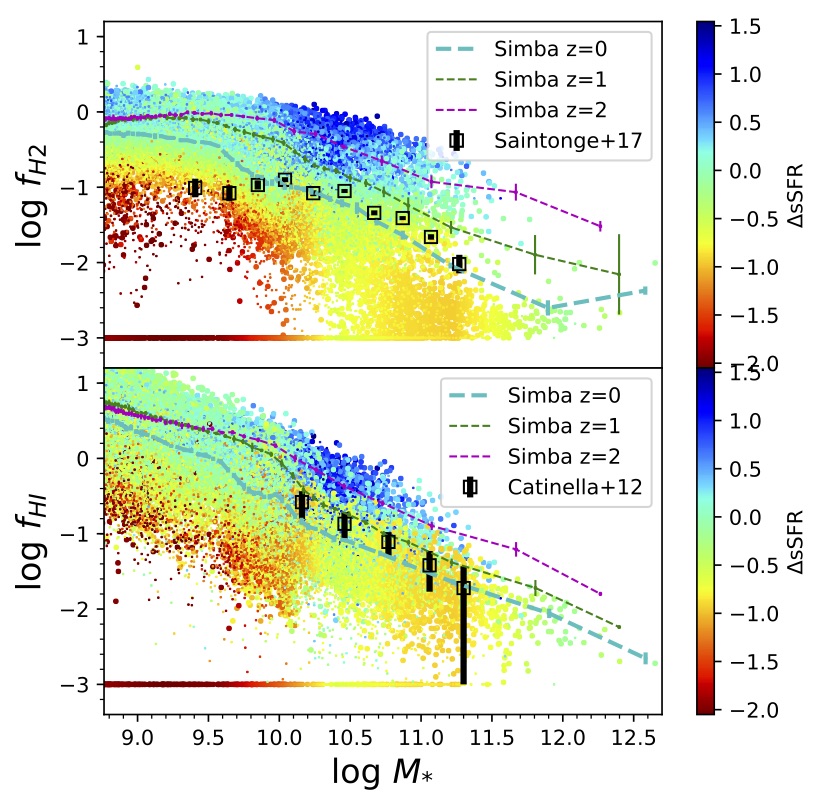}
    \vskip-0.2in
    \caption{Molecular (top) and neutral (bottom) gas fractions $M_{H2}/M_*$ and $M_{HI}/M_*$ as a function of $M_*$.  The points show $z=0$ values from \simba\ colour-coded by the deviation in sSFR from the star-forming main sequence -- bluer points have higher-than typical SFR, redder have lower.  A running median at $z=0$ is shown as the cyan dashed line.  For comparison we show the running medians at $z=1,2$ (green, magneta lines).  Observations of $f_{H2}$ from xCOLDGASS~\citep{Saintonge:2017} are shown in the top panel, and observations of $f_{HI}$ from GASS~\citep{Catinella:2012} are shown in the bottom panel.  \simba\ predicts gas fraction scalings in good agreement with data, and predicts a small but significant amount of gas even in the most massive quenched systems.}
    \label{fig:fgas}
\end{figure}

\simba\ tracks the neutral (\HI) and molecular (H$_2$) hydrogen separately during its evolution, via sub-grid prescriptions to account for molecular gas production and destruction, and approximate self-shielding that results in neutral gas.  Thus \simba\ lends itself to testing against a complementary set of constraints: the scaling relations of \HI\ and H$_2$ gas fractions versus $M_*$.  Recent millimetre and radio observation data has greatly expanded our knowledge of gas contents for low-$z$ galaxies, with constraints at higher $z$ promising continued rapid advancement in the near future.  

Figure~\ref{fig:fgas} shows the scaling relations for H$_2$ (top) and \HI\ (bottom) mass fractions, versus $M_*$.  The points show individual galaxies at $z=0$ colour-coded by their sSFR deviation from the main sequence at that $M_*$ ($\Delta$sSFR), where the main sequence is defined by fitting a running median to the main sequence. The running mean of the gas fractions is shown as the cyan dashed line.
Observations are shown from the mass-selected GASS \HI\ survey~\citep{Catinella:2012} and its follow-up COLDGASS survey~\citep{Saintonge:2017} that obtained H$_2$ masses from CO emission measurements.  We further show the mean predicted trends at $z=1$ (green dashed) and $z=2$ (magenta dashed), to illustrate how these quantities evolve.

Overall, \simba\ does an excellent job of reproducing the trends in both molecular and neutral gas fractions with stellar mass.  There is a hint that the amplitude of both is low by about 0.1~dex, but given observational uncertainties such as the CO-to-H$_2$ conversion factor which is poorly determined particularly in the low-mass ($M_*<10^{10}M_\odot$) regime \citep[e.g.][]{narayanan12a,bolatto13a}, as well as theoretical uncertainties in the approximate way that self-shielding is applied, galaxy gas contents can be considered to be a remarkably successful prediction of \simba.

We note that our massive galaxies have a non-trivial amount of cold gas, despite their very low sSFR.  This was not the case in \mufasa~\citep{Dave:2017a}, where the most massive galaxies were devoid of essentially any cold gas. Recent observations seem to suggest that, perhaps surprisingly, many massive quenched galaxies contain substantial cold gas fractions of up to a percent or more~\citep[e.g.][]{Young:2014}, which is consistent with \simba's predictions. The efficiency of star formation from the molecular gas is, however, low. \simba\ qualitatively reproduces this trend, possibly because the cold gas generally sits in a more extended configuration where the densities are not as high.  Since the star formation rate is proportional to $\rho^{1.5}$, this means that even if the gas has high molecular content, its low density will curtail star formation relative to the same gas being in a compact configuration.  The origin and fate of this cold dense gas is unclear; it could be a transient phase brought in by satellites, or else a stable phase maintained in a more diffuse configuration owing to the presence of hot gas and AGN feedback.  We will examine the exact nature of cold gas in quenched galaxies in future work.

Finally, the colours of the \simba\ points in Figure~\ref{fig:fgas} indicate the deviation of a given galaxy from the main sequence.  There is a clear trend that galaxies that are more gas-rich at a given $M_*$ have a higher SFR.  This is unsurprising for H$_2$ in our models, given that star formation is tied to the molecular content.  It is somewhat more surprising to see this for the \HI\ fraction, but such a correlation was also seen fairly strongly in \mufasa~\citep{Dave:2017a}, though not as strong as for H$_2$~\citep{Rafieferantsoa:2019}.  Such trends have also been noted in observations~\citep{Bothwell:2013} and in EAGLE~\citep{Lagos:2016}.

\subsection{Gas-phase and stellar mass-metallicity relations}

\begin{figure*}
    \centering
	\subfloat{\includegraphics[width=0.5\textwidth]{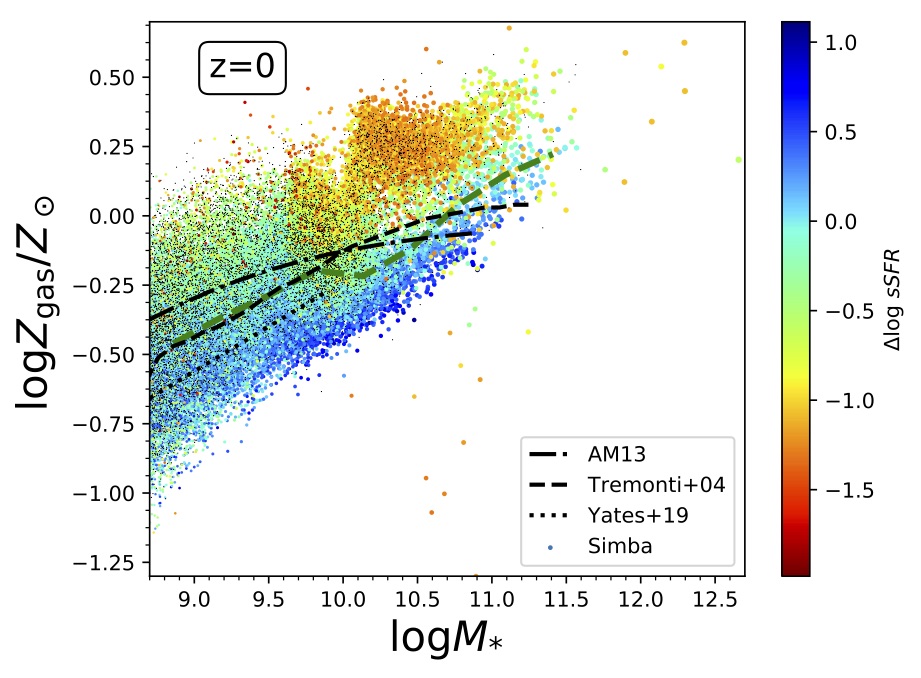}}
    \subfloat{\includegraphics[width=0.5\textwidth]{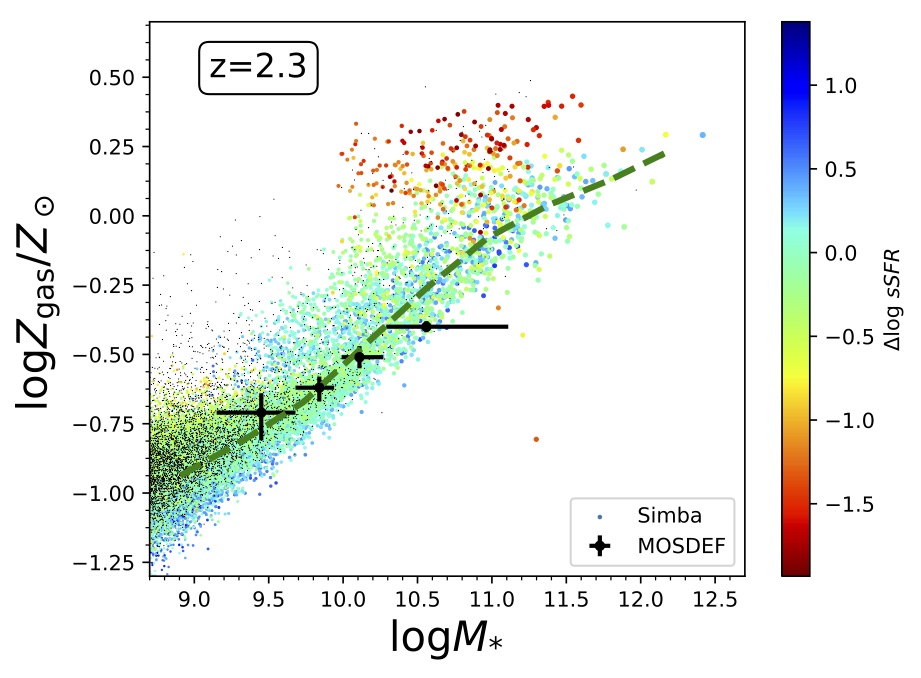}}
    \hfill
    \vskip-0.1in
    \caption{Gas-phase mass-metallicity relation at $z=0$ (left) and $z=2.3$ (right) from \simba.  Points are colour-coded by deviation in sSFR from the star-forming main sequence.  Running median values are shown as the green lines.  At low-$z$, best-fits to observations from \citet{Tremonti:2004,Andrews:2013,Yates:2019} are shown as the black lines.  At $z\approx 2.3$, observations are shown from the MOSDEF survey~\citep{Sanders:2015}.  \simba\ reproduces the gas-phase MZR well at both redshifts, with a noticeable second-parameter dependence on SFR particularly at low-$z$.}
    \label{fig:mzr}
\end{figure*}

\begin{figure}
    \centering
	\includegraphics[width=0.5\textwidth]{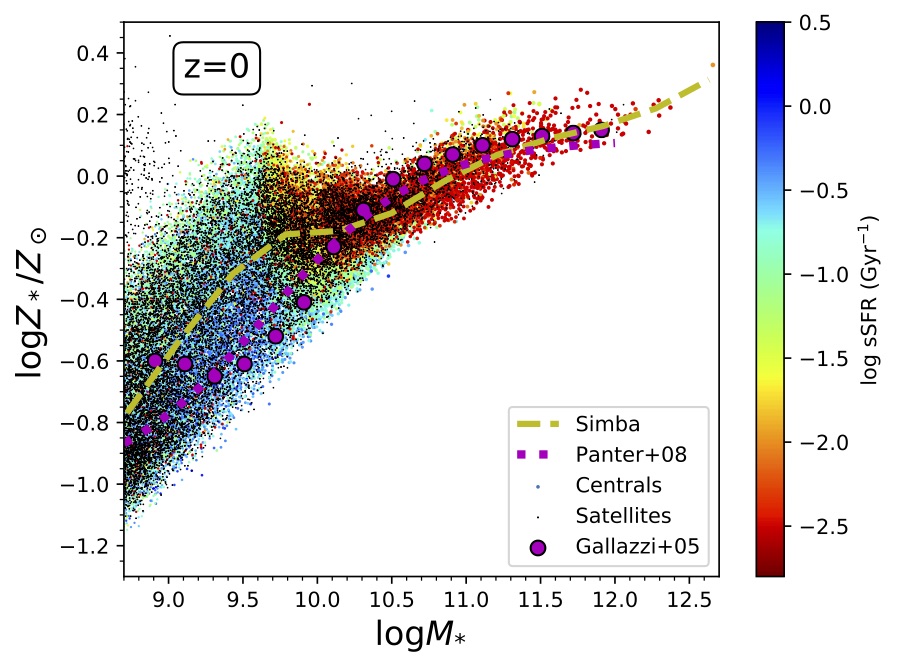}
    \hfill
    \vskip-0.15in
    \caption{Stellar mass--stellar metallicity relation at $z=0$ from \simba, with a running median shown as the dashed yellow line.  Points are colour-coded by specific SFR.  Observations are shown from \citet{Gallazzi:2005,Panter:2008}.  \simba\ reproduces the stellar metallicities of galaxies fairly well, although it appears that low-mass star-forming galaxies tend to have somewhat higher metallicities than typically observed.}
    \label{fig:Zstellar}
\end{figure}

\simba\ tracks the production and distribution of various heavy elements, through several nucleosynthetic channels.  Produced metals can be carried out from galaxies via outflows, which in \simba\ are typically mildly enriched compared to the mean ISM metallicity (see \S\ref{sec:code}).  Additionally, \simba\ locks individual metals into dust, removing them from the gas phase.  Hence predictions for the relationship between galaxy stellar mass and metallicity, which is observed to be among the tightest relations known for galaxies, tests how numerous aspects of \simba\ work together to establish galaxy metallicities. 

Metals can be associated with gas, stars, or dust.  Measurements of the gas-phase metallicity reflect a balance between relatively pristine inflow and ejection of enriched material via outflows, and thus provide a direct constraint on the mass outflow rate in gas-rich (star-forming) galaxies~\citep{Finlator:2008}.  The stellar metallicity is measured from stellar atmospheric absorption lines that reflect the accumulated metals from both gas and dust that ended up locked in the stars.  The inclusion of dust production and destruction model can in principle therefore decouple the stellar and gas phase metallicities.  Here we present predictions for the gas-phase and stellar metallicity scaling relations from \simba.

Figure~\ref{fig:mzr} shows the gas-phase mass-metallicity relation (gMZR) at $z=0$ (left) and $z=2.3$ (right).  The gas-phase metallicity is computed as the SFR-weighted oxygen abundance in all galaxy gas particles, normalized to the solar value of 1.34\% \citep{Asplund:2009}. Points show central galaxies colour-coded by their deviation from the main sequence, as in Figure~\ref{fig:fgas}; black points are satellite galaxies. A running median for star-forming galaxies (sSFR$>10{-1.8+0.3z}$Gyr$^{-1}$) is shown as the dashed green line.  Fits to observations at $z=0$ are shown from strong emission line fitting \citep[black dashed]{Tremonti:2004}, stacked measurement of direct metallicities \citep[black dot-dashed]{Andrews:2013}, and individual semi-direct metallicities~\citep[black dotted]{Yates:2019}.  Observations at $z=2.3$ are shown from the MOSDEF survey~\citep{Sanders:2015}.

\simba\ predicts a gas phase mass-metallicity relation that agrees quite well with observations, lying generally in agreement with the range of current observational determinations.  The metallicities may be slightly too high at the highest masses, but this turns out to be strongly dependent on the assumed cut for star-forming galaxies; a more stringent cut would lower the massive end fit, and highlights the sensitivity of MZR predictions there to precise selection effects.  

\mufasa\ produced a gMZR that was slightly too steep, in addition to having an amplitude that roughly agreed with data only because of an arbitrary halving of the SNII yields~\citep{Dave:2017a}.  In \simba, metals are locked into dust, increasingly so at higher metallicities and hence larger masses.  This likely leads to a suppression of the gas-phase metallicity in massive galaxies and thus a flatter gMZR, as well as a lower amplitude.  We will examine the impact of dust on the metal content of galaxies in more detail in future work (Li et al., in prep.).

Since $z=2.3$, \simba\ produces more metal evolution at the low-mass end, in general agreement with observations suggesting that the most metal-rich galaxies are in place at early epochs~\citep[e.g.][]{Zahid:2014}.  Star-forming galaxy metallicities are in good agreement with the MOSDEF data, suggesting that the amount of metal evolution from $z\sim 2\rightarrow 0$ is approximately correct in \simba.

\simba\ also shows a clear second parameter dependence on the specific SFR such that at a given $M_*$, galaxies with lower sSFR tend to have higher metallicities. This has been noted observationally as the Fundamental Metallicity Relation~\citep[FMR;][]{Mannucci:2010,Lara-Lopez:2010}.  The existence of the FMR remains somewhat controversial~\citep{Salim:2014,Sanchez:2017,Cresci:2018}, but a careful analysis of the MOSDEF data has revealed such a trend at $z\sim 2$~\citep{Sanders:2018}.  The trend is quite obvious at $z=0$ with the bluest galaxies clearly having the lowest metallicities, but is not quite so evident at $z=2.3$ except at the massive end where a population of quenched galaxies has appeared.  Finally, the small black points showing the satellites tend to lie above the mean relation, in qualitative agreement with data~\citep{Pasquali:2012}.

Figure~\ref{fig:Zstellar} shows the mass-weighted stellar metallicity as a function of $M_*$ at $z=0$.  Points show centrals colour-coded by sSFR, with satellites in black. The yellow dashed line shows a running median.  Observations are shown from \citet{Gallazzi:2005} and \citet{Panter:2008}.

\simba\ nicely reproduces the stellar MZR for massive, quenched galaxies.  At lower masses, the star-forming population dominates, and these tend to have a stellar MZR that is typically slightly higher, with larger scatter, than expected from an extrapolation of the massive galaxy relation.  This owes to the fact that these galaxies have continued to form stars after their more massive counterparts have quenched.  However, no such feature is evident in the observations, and hence \simba\ produces a low-mass stellar MZR that is somewhat too high compared to observations.

In summary, \simba\ does a reasonable job reproducing observed galaxy metallicities, both stellar and gas phase, and the evolution out to Cosmic Noon.  The fact that no arbitrary normalisation was required as in \mufasa\ is a step forward, suggesting that \simba\ is locking metals into dust in a realistic manner; we will explore this more in \S\ref{sec:dustprop}.

%\subsection{Galaxy colour--mass relation}

%A key reason for the inclusion of AGN feedback is to reproduce a tight red sequence as observed, with particularly the most massive galaxies being old, forming stars very slowly, and metal-rich.  The tightness of the red sequence suggests very little ongoing star formation in such massive galaxies, something that has been a challenge to achieve in self-consistent models of AGN feedback.  Hence the red sequence, here explored as the related colour-mass relation, represents a key test of the efficacy of galaxy quenching.

\subsection{Galaxy photometric sizes}

\begin{figure}
	\includegraphics[width=0.5\textwidth]{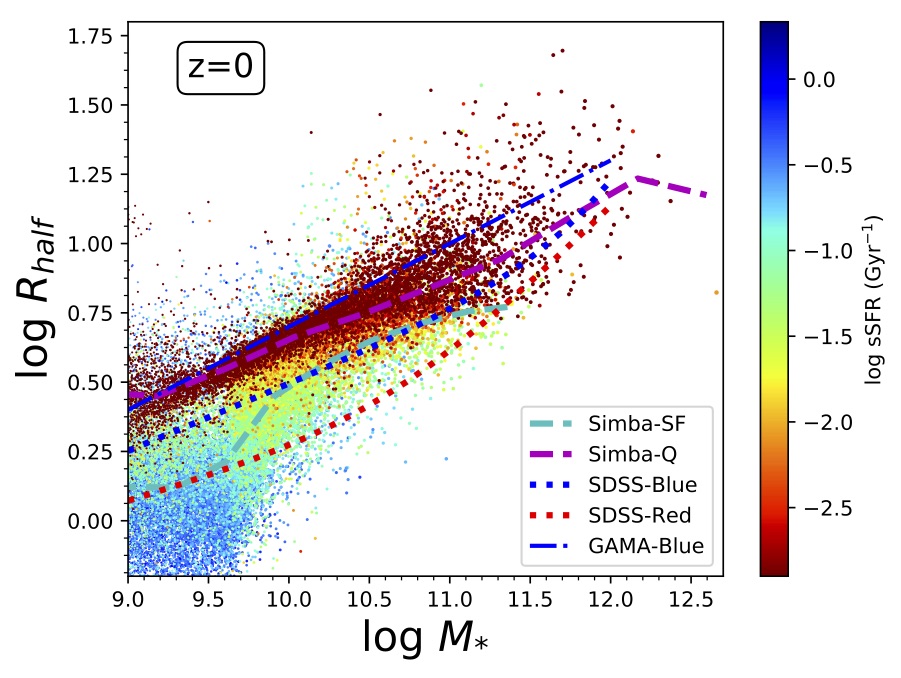}
    \vskip-0.1in
    \caption{$R$-band 2-D projected half-light radii of \simba\ galaxies at $z=0$, as a function of $M_*$.  Points are colour-coded by sSFR.  We fit median relations to the red and blue sub-samples, delineated by $10^{-1.8}$~Gyr$^{-1}$, as the cyan and magenta dashed lines lines, respectively. Observational relations are shown from \citet{Zhang:2017} from SDSS, split into red and blue galaxies.  \simba\ broadly reproduces the sizes of star-forming galaxies, but fails to show the observed trend that quiescent galaxies have a much steeper slope and are much more compact at low masses.    }
    \label{fig:halfmass}
\end{figure}

Modern cosmological simulations typically have sufficent resolution to resolve the size of galaxies, even if the detailed structure of the ISM remains unresolved.  Illustris highlighted the ability for simulations to produce galaxies populating the full range of the Hubble sequence~\citep{Vogelsberger:2014}.  For EAGLE, galaxy sizes provided a key constraint on their star formation feedback implementation~\citep{Schaye:2015}, namely that they employ a steeper dependence of the star formation rate on the density in dense gas in order to prevent galaxies from being overly compact.  \simba\ did not use sizes to tune the feedback model, so instead they provide a test of it.

To conduct a fair comparison to observed sizes, we compute projected galaxy half-{\it light} radii in the $R$-band. We obtain $R$-band luminosities from the \citet{Bruzual:2003} models interpolated to the age and metallicity of each star particle. The radius is determined for each galaxy by averaging 2-D half-light projections along the ${x,y,z}$ axes.  Figure~\ref{fig:halfmass} shows the galaxy half-light sizes at $z=0$ from \simba\ versus stellar mass.  We colour-code the points by sSFR as before, and fit separate running medians for quenched (\simba-Q) and star-forming (\simba-SF) populations (magenta and cyan lines), where we divide the two populations at sSFR$=10^{-1.8}$~Gyr$^{-1}$ as before. For comparison we show observations from SDSS~\citep{Zhang:2017}.  The SDSS sample has been subdivided into red and blue galaxies, albeit with a criterion based on photometry, not sSFR.

Star-forming galaxy sizes in \simba\ show an amplitude and scaling with $M_*$ that agrees quite well with the observed slope, which is encouraging.  There is a suggestion that low-mass galaxies are too small, but this occurs in the mass range where the number of particles is below a few hundred, and given that the sizes are light-weighted, stochasticity can give rise to smaller-than-expected sizes.  We note that a stellar mass-weighted size does not show this drop-off at low masses.  But for well-resolved star-forming galaxies, the sizes are in quite good agreement with data.  This is an important success that did not require any specific tuning of the feedback model.

In contrast to the star-forming systems, \simba\ shows quenched galaxy sizes that are quite discrepant with observations.  Massive galaxy sizes are in reasonable agreement with data, but the lowest-mass quiescent galaxies are up to $\sim\times 3$ larger than the comparable sample in SDSS, showing that the size--mass trend for passive galaxies is incorrect in \simba; indeed, in \simba\ the low-mass passive galaxies are actually larger than the star-forming ones, which is opposite to the observed trend.  There are a number of potential reasons for this.  The large number of stellar orbits in older quiescent galaxies tends to puff out the distribution numerically. The discrepancy could also represent a failing in physics, if for instance low-mass galaxies are preferentially quenched via some rapid mode such as merging, violent disk instability, or stripping that is simultaneously associated with compactification~\citep[e.g.][]{Tacchella:2016}.  {\color{black} Alternatively, it could be a failing of the feedback physics associated with quenching low-mass galaxies.}  We can test this issue directly with higher resolution runs once they complete. For now, we note that the sizes of small quenched galaxies in \simba\ is a clear failing of the current model, in contrast to its success at reproducing the sizes of star-forming galaxies.

\subsection{Halo gas fractions}

\begin{figure}
	\includegraphics[width=0.45\textwidth]{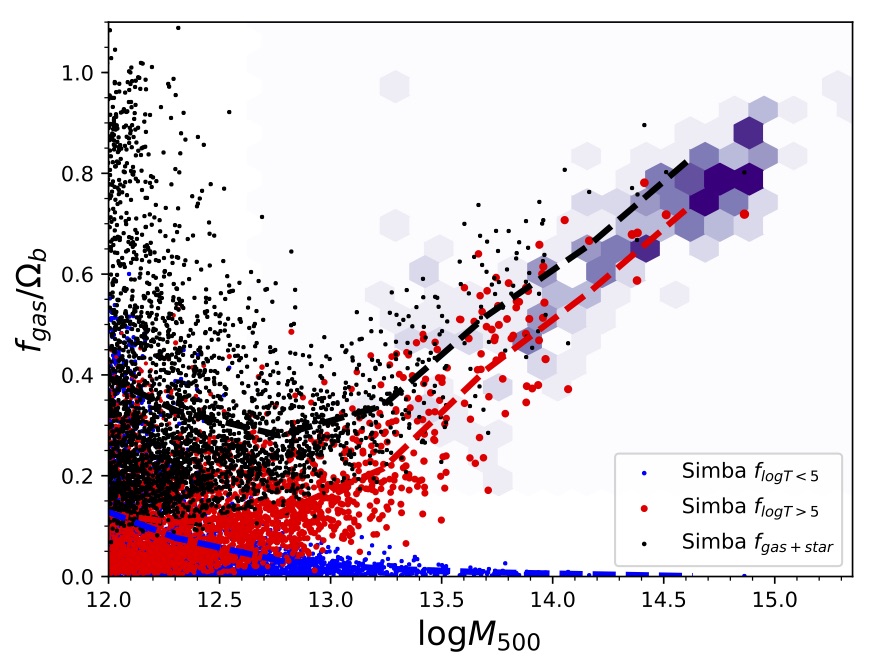}
    \vskip-0.1in
    \caption{Gas fractions as a function of $M_{500}$ at $z=0$.  Black points show the total baryon fraction within the halo, and red and blue points show the gas fractions subdivided at $10^5$K. Purple hexbins show an observational compilation of hot gas fractions from \citet{McCarthy:2017}, to be compared to the red points. \simba\ does a reasonable job of reproducing the observed trend of hot gas fraction with $M_{500}$, which has been difficult for previous simulations to achieve without tuning.}
    \label{fig:hotgasfrac}
\end{figure}

In \simba, AGN feedback provides the primary energy input that serves to quench massive galaxies in large halos.  Such energy input can  concurrently have a strong impact on the amount and thermal state of hot gas within those halos. In particular, it can evacuate gas even from fairly sizeable halos, somewhat by entraining gas in jets but mostly by depositing heat that results in the gas becoming unbound to the halo.  These processes result in halo gas fractions that deviate strongly from the mean cosmic baryon fraction, a departure that can be measured in real systems via X-ray emission from intra-group and intra-cluster gas. Such observations thus provide an important constraint on the AGN feedback model.

The hot gas fraction as a function of halo mass has been a challenging constraint for modern cosmological AGN feedback models to reproduce~\citep{McCarthy:2017}.  In Illustris, it was found that the AGN feedback mechanism over-evacuated hot gas from group-sized halos compared to observations~\citep{Genel:2014}, which provided one motivation for the new AGN feedback model in Illustris-TNG~\citep{Weinberger:2018}. {\color{black} Nonetheless, while closer than Illustris, TNG somewhat overpredicts the observed hot gas fractions~\citep{Barnes:2018}.  EAGLE likewise overpredicts the hot gas fractions, while the
{\sc Bahamas} simulation suite was able to match this with mild tuning ~\citep{McCarthy:2017}.}  Here we examine this constraint for \simba.

Figure~\ref{fig:hotgasfrac} shows the baryon fractions as a function of halo mass ($M_{500}$) from \simba.  $M_{500}$ is computed as the radius enclosing 500 times the critical density, centered on the most bound halo particle.  Black points show the total baryon fraction, red points show hot gas ($T>10^5$K) fractions, and blue show cold gas ($T<10^5$K).  Colour-coordinated lines show the running median values.  Note that the black points include the stellar (and black hole) contribution, which is not explicitly shown.  A compilation of observations from \citet{McCarthy:2017} is shown as the purple hexbins.  All fractions have been scaled to $\Omega_b=0.048$, so a halo at unity has its cosmic share of baryons.

At $10^{12}M_\odot$ halos have about 40\% of the cosmic baryon fraction within $R_{500}$, with a large scatter.  This dips to $\sim 30\%$ at $10^{12.5-13}M_\odot$, before rising again to large halo masses.  The dip owes to jet AGN feedback, which we have checked by comparing to the No-jet test run, which shows baryon fractions around 90\% for all halos over this mass range (and a much flatter trend of hot gas fraction versus halo mass).  This shows that the energy input required to quench galaxies can cause substantial evacuation of group-sized halos, as has been noted in e.g. Illustris~\citep{Genel:2014}. This strong evacuation has important implications for using these systems as probes of cosmology, which we will probe in future work.

The hot baryon fraction is shown in red, which can be compared to the observations shown in purple.  In massive systems, the total baryons are dominated by this hot phase.  Most of this hot gas is near the virial temperature, so the results are insensitive to the exact value of the cut at $10^5$K.  Comparing to observations, we see that \simba's halos have hot baryon fractions that are well in agreement with data in the overlapping mass range, in both amplitude and scatter. We note that there was no tuning done to obtain this agreement.  The halo hot baryon fraction is thus a non-trivial success of \simba's AGN feedback model, and shows that \simba\ evacuates halo baryons in a manner that is concordant with observations. In Borrow et al. (in prep.) we will examine quantitatively where these evacuated baryons end up.

\subsection{Black hole mass vs stellar mass}

\begin{figure}
	\includegraphics[width=0.5\textwidth]{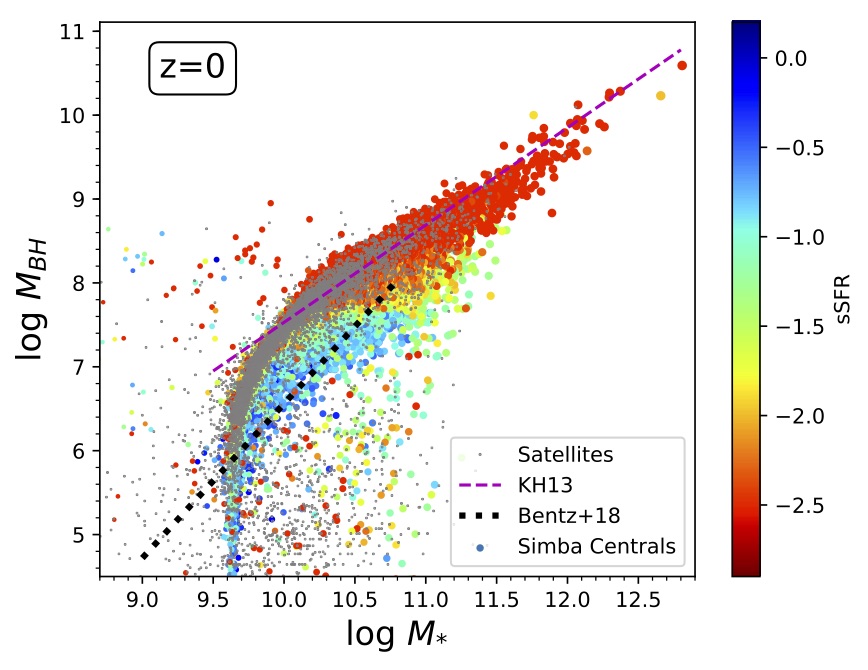}
	\vskip-0.1in
    \caption{$\mbh-M_*$ relation at $z=0$ in \simba.  Points show galaxies, with centrals colour-coded by specific SFR, and satellites as the grey points.  Observations are shown from \citet{Kormendy:2013} for comparison with bulge-dominated (redder) systems, while \citet{Bentz:2018} shows the relationship more appropriate for spiral star-forming systems at lower $M_*$.  \simba\ broadly reproduces these observed relations in its appropriate galaxy populations.}
    \label{fig:mbhms}
\end{figure}

The canonical relation that highlights the connection between galaxies and their central supermassive black holes is the relationship between the black hole mass and the galaxy bulge mass or stellar velocity dispersion~\citep[e.g.][]{Magorrian:1998,Kormendy:2013,McConnell:2013,Graham:2016,Bentz:2018}.  Modern galaxy formation models that track black holes typically have free parameter(s) that are tuned to match these relations; in \simba, this is set by the accretion efficiency tuned to $\epsilon_m=10$\%, while most other cosmological simulations (based on Bondi accretion) tune the AGN feedback efficiency.  In previous works, \citet{Angles:2015} and \citet{Angles:2017} showed that the $\mbh-M_*$ relation emerged naturally from the torque-limited accretion model, without or with AGN feedback, respectively.  But these studies were done via post-processing or without star formation feedback.  Here we examine whether the full physics model in \simba\ likewise reproduces the relationship between black hole mass and galaxy properties.  

Figure~\ref{fig:mbhms} shows the black hole mass--stellar mass relation at $z=0$ for \simba\ galaxies. Central galaxies are shown colour-coded by specific SFR, while satellite galaxies are indicated by grey points.  The relationship for galaxy bulges is shown from \citet{Kormendy:2013} as the magenta dashed line; this is an appropriate comparison sample for bulge-dominated galaxies, which are expected to be the quiescent systems with redder points.  Meanwhile, \citet{Bentz:2018} assembled a sample of reverberation-mapped galaxies, and found the steeper relation shown as the blue dotted line. In their case, the lower-mass systems are predominantly late-type galaxies, while the most massive systems are early-type.  Hence in the region plotted, the \citet{Bentz:2018} sample is probably best compared to later-type systems, and therefore star forming (bluer points).  We note that all observational relations show a large scatter, typically at least 0.3~dex, which is not represented on this plot.

\simba\ black holes generally lie in the range of observations.  Although we tuned $\epsilon_m=0.1$ in order to obtain the correct amplitude of the relation, the slope of the relation is not tunable, particularly in terms of different galaxy sub-samples. Hence the agreement of the quiescent galaxy slope with the bulge-dominated galaxy black holes, and likewise the general agreement of the star-forming galaxies with the lower black hole masses at a given $M_*$, is in good agreement with observations.  We note that there is some disagreement on whether the late-type galaxies have a steeper slope or the same slope but offset to lower black hole masses~\citep[e.g.][]{Graham:2016,Savorgnan:2016,Bentz:2018}, but \simba\ predictions are broadly compatible with either scenario.

At the low-mass end, consistent with \citet{Angles:2015}, torque-limited accretion grows black holes very quickly once the galaxy stellar mass exceeds $3\times 10^9M_\odot$, which is where we choose to seed the black holes in \simba.  Hence the rapid rise is not directly physical but a numerical artifact of our seeding prescription, though it is intended to mimic the physical effect of black hole growth suppression due to early star formation seen in e.g. FIRE \citep{Angles:2017c}.  Also, we note that we attempt to keep black holes fixed to the centre of the galaxy potential well, but in dense regions this does not always work owing to the shallow potential wells in poorly resolved galaxies, so black holes can move between galaxies and thus merge. We continue to test for approaches for better handling this given the poor cosmological resolution.

Overall, \simba\ predicts a relationship between black hole and stellar masses in agreement with observations.  In upcoming work we will examine black hole scaling relations in more detail, but for now, the good agreement corroborates the idea that black holes in \simba\ grow in accord with observations, and thus the feedback energy released by black holes and used by \simba\ to quench galaxies is plausible.

\subsection{Dust Properties}\label{sec:dustprop}

\begin{figure}
	\includegraphics[width=0.5\textwidth]{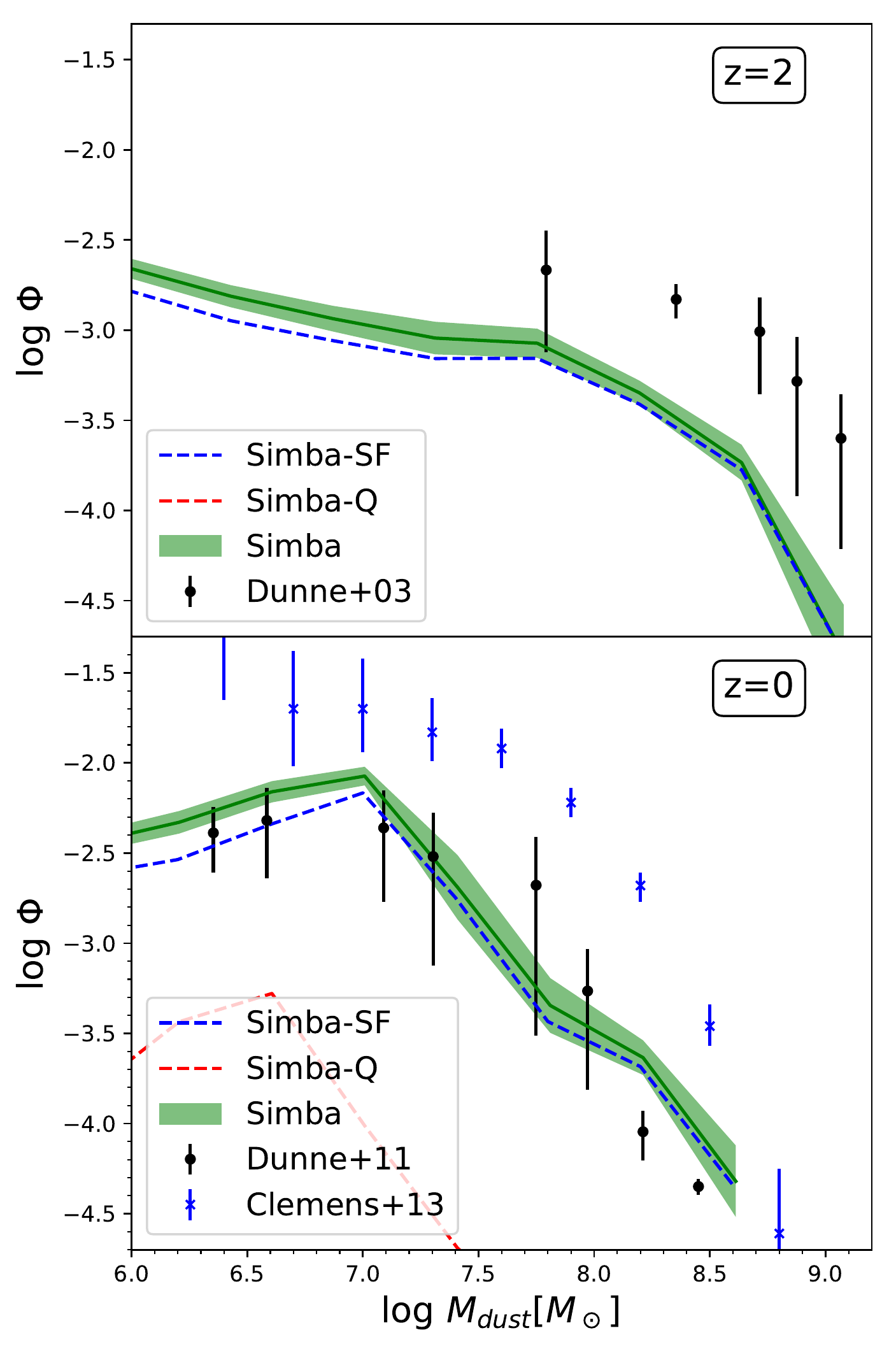}
    \vskip-0.1in
    \caption{Dust mass function from \simba\ at $z=0$, shown as the green shaded region. DMF is split into star-forming and quenched samples, shown as blue and red dashed lines, respectively.  Observations are shown from \citet{Dunne:2011} and \citet{Clemens:2013}; differences owe to assumptions regarding inferring dust mass from far-IR flux.  \simba\ reproduces the observed shape of the DMF, and is in the range of observed amplitudes, modulo systematic uncertainties.}
    \label{fig:mfdust}
\end{figure}

\begin{figure}
	\includegraphics[width=0.48\textwidth]{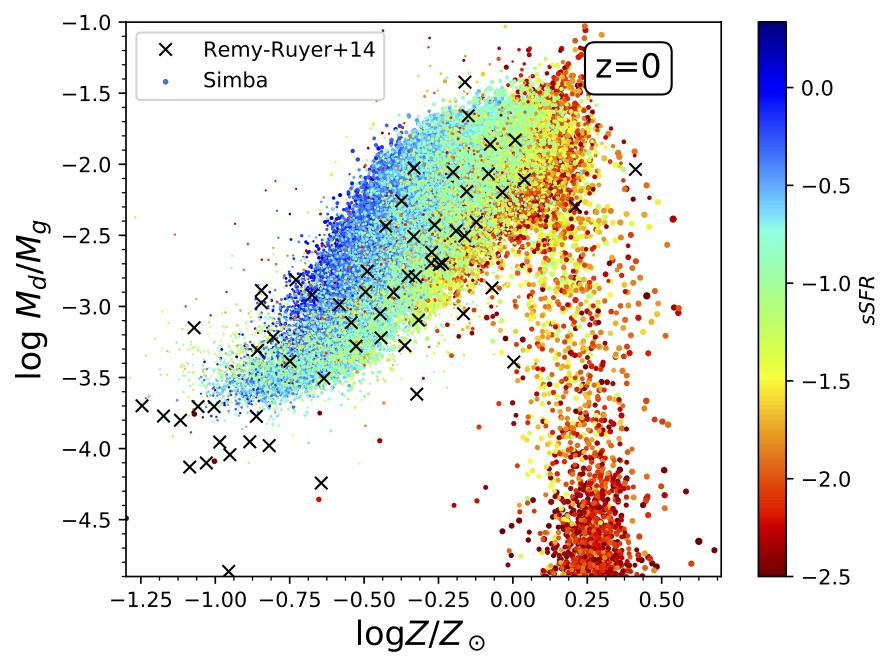}
    \vskip-0.1in
    \caption{Dust to gas ratio as a function of gas-phase  metallicity at $z=0$ in \simba\ galaxies, colour-coded by specific SFR.  Observations are shown as crosses from \citet{Remy-Ruyer:2014}.  \simba\ reproduces the observed trend and amplitude in dust-to-gas ratios.}
    \label{fig:dusttogas}
\end{figure}

\begin{figure}
	\includegraphics[width=0.48\textwidth]{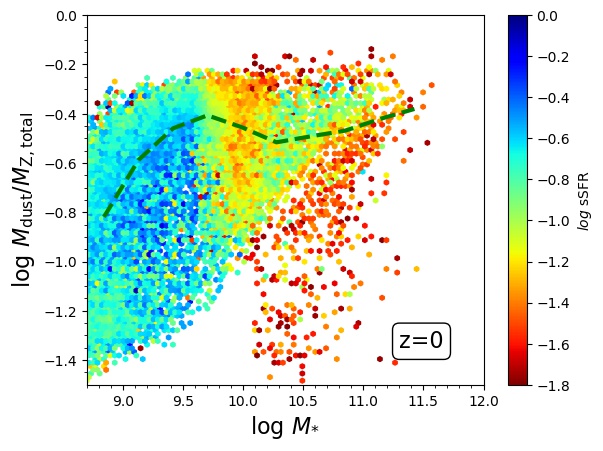}
    \vskip-0.1in
    \caption{Metal mass fraction locked in dust, as a function of $M_*$, for star-forming galaxies in \simba.  Plot is colour-coded by the mean sSFR within each hexbin.  Except for the smallest galaxies, typically one-third of the metals are locked in dust.}
    \label{fig:dtm}
\end{figure}

\simba\ includes a model to form and destroy dust from metals within the ISM of galaxies during the simulation run.  As a basic check on the production of dust, here we examine two measurables tracking dust in galaxies: The dust mass function, and the dust-to-gas ratio.

Figure~\ref{fig:mfdust} shows the $z=0$ (bottom panel) and $z=2$ (top) dust mass function (DMF) from \simba\ (green line), versus two $z=0$ observational determinations from \citet{Dunne:2011} and \citet{Clemens:2013}, and a $z=2$ determination from \citet{Dunne:2003}.  At $z=0$, \simba\ agrees well with \citet{Dunne:2011}, but not \citet{Clemens:2013}.  The difference between the two can be traced to their assumption of the dust mass opacity coefficient used to infer the dust mass from far-IR data; \citet{Clemens:2013} showed that under the same assumption of this quantity, the two results agree.  Hence given current uncertainties in inferring dust masses, it is probably premature to use the DMF as a strong constraint on models.  But \simba's DMF is at least within the ballpark of currently observed values, with good agreement in the overall DMF shape.  {\color{black} Unsurprisingly, the DMF is dominated by star-forming galaxies (blue dashed line), which is as observed~\citep{Beeston:2018}.

The $z=2$ DMF is compared to observations from \citet{Dunne:2003}, and shows a deficit of $\sim\times 3$ in the number density of galaxies at a given dust mass. {\color{black}We note that the observational DMF by \cite{Dunne:2003} is from surveys of sub-mm sources with large beam sizes, which could result in multiple objects being blended within one beam therefore overestimating their dust masses.} If one regards the \citet{Clemens:2013} results at $z=0$ to be more accurate, as confirmed by \citet{Beeston:2018}, then the shortfall in the predicted DMF is very similar at both redshifts.  This suggests that the evolution in dust masses in \simba\ is viable, but the overall dust production is short, or else destruction is too efficient.  It may be possible to remedy this with differing choices of dust parameters; we are exploring this.  We note that our $z=0$ DMF agrees well with the predictions from cosmological simulations of \citet{McKinnon:2017}, owing in part to tuning of each model, but our $z=2$ DMF is significantly higher than theirs.}

Figure~\ref{fig:dusttogas} shows the $z=0$ dust-to-gas ratio (DGR) as a function of gas-phase metal abundance. Points are colour-coded by sSFR.  \simba\ is in good agreement with the data shown \citep[crosses]{Remy-Ruyer:2014}, showing a slope of increasing DGR with metallicity as observed.  In massive quenched systems, the DGR drops quickly.  

{\color{black}Finally, Figure~\ref{fig:dtm} shows the fraction of metals locked into dust at $z=0$ for star-forming galaxies. The green line shows the running median.  For galaxies with $M_*\ga 10^{9.5}M_\odot$, the fraction is typically 30-40\%, but drops significantly to lower masses.  This mostly explains why the mass-metallicity relation agrees with observations in \simba\ without the {\it ad hoc} reduction of the yields by $\times 2$ as in \mufasa.
Low-sSFR galaxies also have fewer metals locked into dust, as AGN feedback returns metals locked in dust into the gas phase.}  We will examine galaxy dust content and evolution in significantly more detail in forthcoming work (Li et al., in prep.), but these preliminary comparisons suggest that \simba's dust tracking model yields plausible galaxy dust contents.

\section{AGN feedback variations}\label{sec:variants}

AGN feedback is believed to be responsible for quenching galaxies.  \simba\ includes three different forms of AGN feedback:  Radiative mode AGN winds, AGN jets, and X-ray feedback.  In this section we examine the importance of these various modules in producing a quenched galaxy population, by running simulations with AGN jets and X-ray feedback off, and with only X-ray feedback off.  We always include radiative AGN winds.  For these tests we use $50\hmpc$, $2\times 512^3$ simulations run to $z=0$, with \simba\ input physics and parameters except for the AGN feedback variations.

\begin{figure}
	\includegraphics[width=0.45\textwidth]{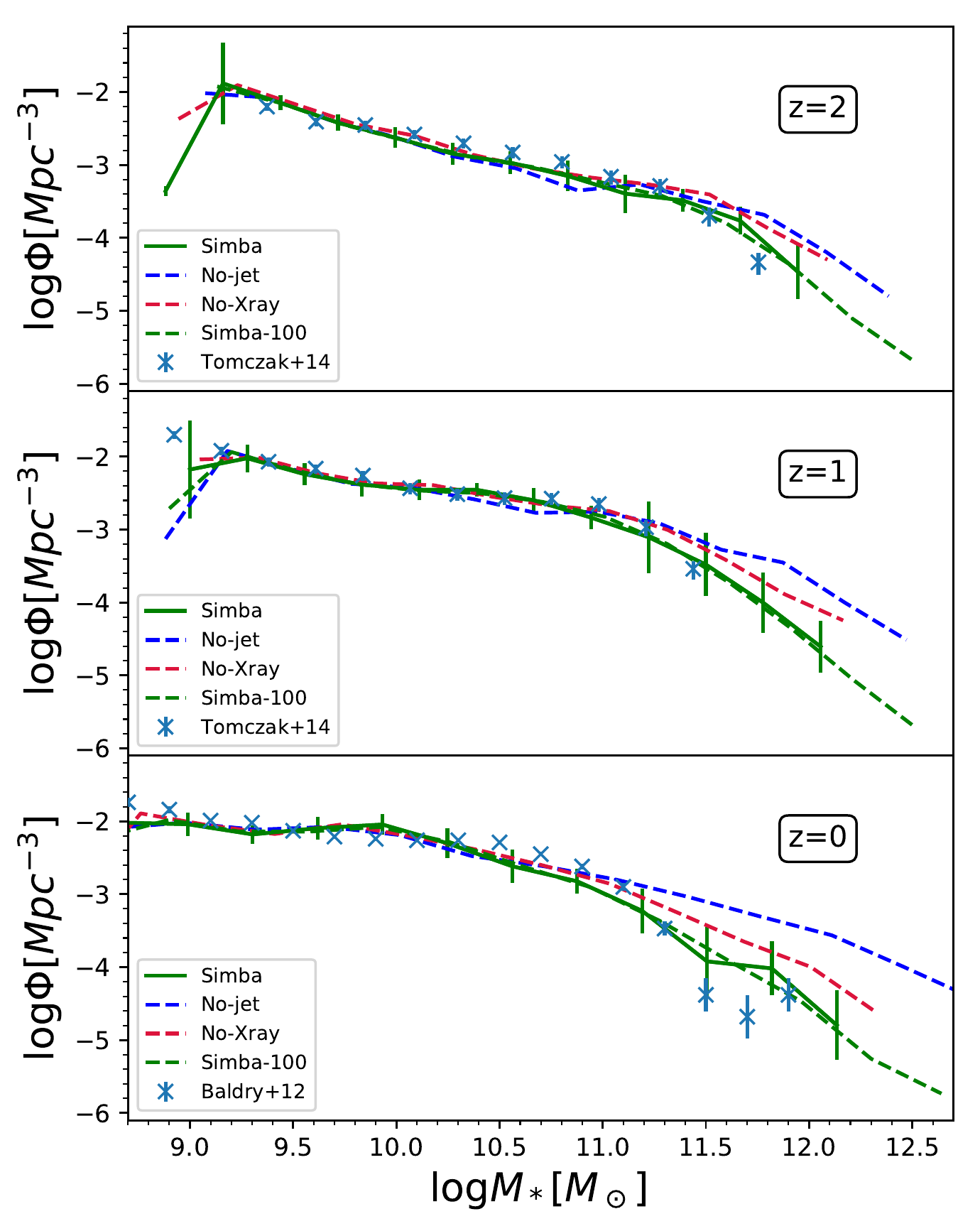}
    \caption{Stellar mass function evolution from $z=2\rightarrow 0$ in $50\hmpc$, $512^3$ test runs with different AGN feedback variants:  Original \simba\ (green solid), Jet and X-ray feedback both turned off (No-jet; blue dashed), only X-ray feedback turned off (No-Xray; red dashed).  For comparison we show the main $100\hmpc$ \simba\ run (green dashed) reproduced from Figure~\ref{fig:mfwind}, as well as selected observations as indicated.  Turning on just the jet feedback (No-Xray) results in a substantial truncation of the GSMF that does not occur without jets (No-jet).}
    \label{fig:mfvar}
\end{figure}

\begin{figure}
	\includegraphics[width=0.45\textwidth]{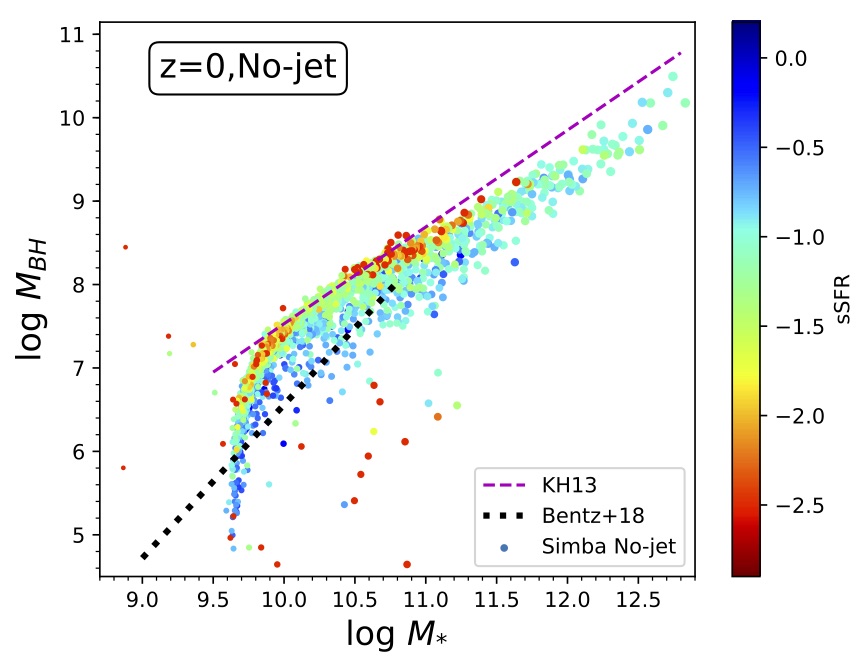}
	\includegraphics[width=0.45\textwidth]{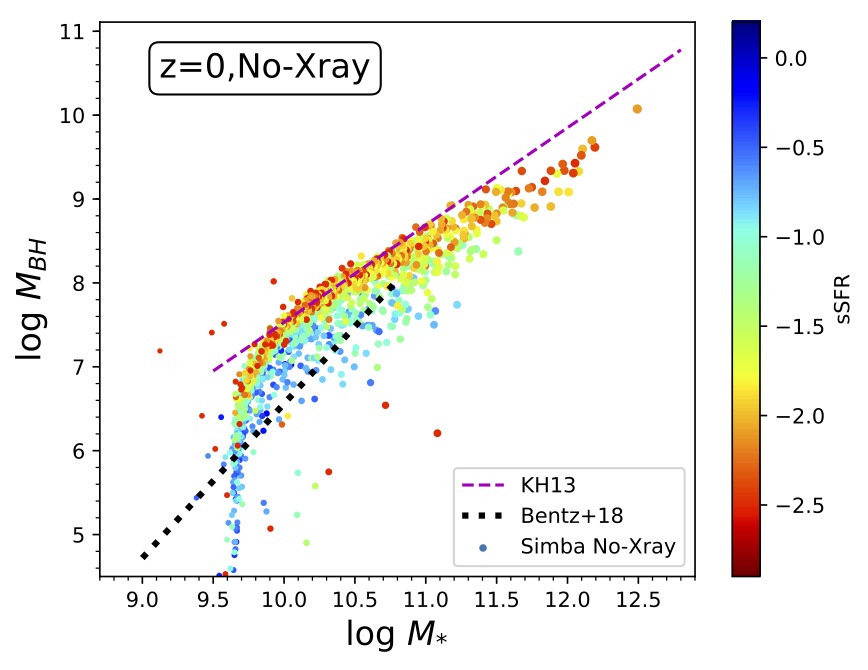}
	\vskip-0.1in
    \caption{$M_{BH}-M_*$ relations in test simulations with jet and X-ray black hole feedback turned off (No-jet, top panel), and jets on but X-ray feedback turned off (No-Xray, bottom).  By comparing to Figure~\ref{fig:mbhms}, the jet feedback is seen to enact most of the quenching, but the X-ray feedback is important for fully quenching the most massive galaxies.}
    \label{fig:mbhmsvar}
\end{figure}
Figure~\ref{fig:mfvar} shows the GSMF for the full \simba\ physics run, a {\it No-X} run turning off only X-ray feedback, and a {\it No-jet} run turning off both jet and X-ray feedback, at $z=2,1,0$.  Observations are overplotted as described in Figure~\ref{fig:mfwind}. We do not show $z\geq 3$ results because these variants' GSMFs are indistinguishable there.

Looking at the $z=0$ panel, without jets, the GSMF is strongly overproduced at high masses.  Turning on jets results in much better agreement with full \simba.  This is even true when including X-ray feedback, which makes only a small change to the GSMF.  The redshift evolution shows that the impact of jets is fairly minor at $z\sim 2$ in terms of the GSMF, only impacting the very largest few galaxies.  The importance of jets in truncating the GSMF grows steadily with time, to where without jets the number density of $M_*=10^{12}M_\odot$ galaxies would be an order of magnitude higher, in strong disagreement with data.  These results clearly show that the main driver in truncating the massive end of the GSMF in \simba\ is AGN jet feedback.

A more detailed view of how AGN feedback impacts both stellar and black hole growth can be obtained by examining the $\mbh-M_*$ relation in these variants, shown in Figure~\ref{fig:mbhmsvar}, with galaxies colour-coded by specific SFR as in Figure~\ref{fig:mbhms}. For clarity we show only central galaxies. 

Comparing the {\it No-Jet} version (top panel) to the original \simba\ in Figure~\ref{fig:mbhms} highlights several key points.  As expected, the sub-$M^\star$ objects show little difference in the trends, in either $\mbh$ or sSFR.  However, for massive galaxies, the full \simba\ run shows significantly lower sSFR and somewhat higher $\mbh$, particularly for the most massive galaxies.  This demonstrates more explicitly that the AGN jet feedback is crucial for quenching galaxies.

Interestingly, the {\it No-jet} run still shows a few quenched galaxies at high $\mbh$ around $M^\star$.  These are clearly correlated with the presence of a massive black hole, and would not occur in a model with no AGN feedback at all.  This arises from the fact that we still have radiative AGN winds in our {\it No-jet} run.  These become effective around $M^\star$ because it is at the corresponding halo mass that a significant hot gaseous halo begins to form~\citep{Keres:2005,Gabor:2012}.  The AGN energy can then be deposited into the hot gas, providing a mechanism for quenching the galaxy by shutting off the fuel supply~\citep{Dekel:2009}.

So why do radiative winds cease to be effective at higher masses?  An examination of the energetics shows the reason.  In \simba's AGN feedback model the momentum input is assumed to be constant, which means that the AGN feedback energy scales as the wind velocity.  Since \simba's black hole accretion rates are a quite weak function of $M_*$~(Thomas et al., in prep) while the number of hot halo baryons is growing, this means that the energy injected per hot halo baryon is dropping with the halo mass.  The logarithmic increase in velocity with $\mbh$ (eq.~\ref{eq:vradiative}) is too slow to compensate for this, so one quickly ends up in a situation where the energy injection is insufficient to keep the hot halo baryons near the virial temperature. What is required is a strong increase in the outflow velocity, and hence energy input, in this halo mass regime.  This is why high-velocity AGN jets are crucial for quenching massive galaxies.

The black hole masses in the {\it No-jet} run are also appear to be significantly lower.  However, this can primarily be explained by the effect that $M_*$ values in this simulation are substantially higher, which moves galaxies leftwards in the $\mbh-M_*$ diagram; the black hole masses themselves are not substantially different. The relative roles of Bondi vs. torque-limited accretion in growing black holes across the full mass range over cosmic time will be examined more fully in a forthcoming paper~(Ang\'es-Alc\'azar et al. in prep.).

The {\it No-Xray} run is fairly similar to the full \simba\ run, but there is a slight if noticeable increase in the sSFR in the massive galaxies.  This is not so much as to contribute significant mass growth, hence the GSMF is only modestly affected, but it is higher than typical observed values for massive red and dead ellipticals.  This suggests that the X-ray feedback is important for fully quenching massive galaxies in accord with observations, even if it does not play a leading role in regulating mass growth.

\section{Summary}\label{sec:summary}

We have introduced the new \simba\ suite of cosmological galaxy formation simulations, and explored predictions from a 100 Mpc/h box run with $1024^3$ dark matter and $1024^3$ gas elements.  The most novel aspect of \simba\ is its implementation of black hole growth via torque-limited accretion, and two-mode black hole feedback via bipolar kinetic outflows.  \simba\ further includes numerous updates over its predecessor simulation \mufasa, including a dust production and destruction model.  In this paper we present comparisons to a range of different observational probes measuring the stellar mass, star formation rate, neutral and molecular gas, black hole, and dust properties in \simba.  We show that, in all cases, \simba\ produces galaxies that are in quite reasonable agreement with observations.  While our feedback parameters were generally chosen to follow observations or expectation from high-resolution simulations, some of these observations were used to further tune these parameters.  However, many of them were not, and these represent model predictions that demonstrate the viability of \simba\ as a platform for studying cosmological-scale galaxy evolution.

Here are our main findings:
\begin{itemize}
    \item \simba\ produces a stellar mass function evolution that is in very good agreement with data across all masses at all cosmic epochs, although it may overproduce the massive end slightly by $z=0$.  Quenched galaxies grow substantial in numbers at $z\la 2$, and by $z=0$ they dominate at $M_*>2\times 10^{10}M_\odot$.
    \item \simba's star forming main sequence is in good agreement with observations at $z=0$, and low at $z\approx 2$ by only a factor of two which is explainable via observational systematics.  {\color{black} Predicted quenched fractions at $z=0$ as a function of $M_*$ are in good agreement with observations. } Galaxies at a given $M_*$ that quench first in \simba\ have preferentially larger black holes.
    \item \simba\ gas fractions, both neutral and molecular, show a dropping trend with $M_*$ that is in good agreement with observations.  Gas fractions evolve downwards with time, but even at $z=0$, massive quenched galaxies still typically have some cold gas.
    \item The gas-phase and stellar mass--metallicity relations generally agree with observations at $z=0$ and $z\sim 2$.  The MZR evolves upwards by a factor of $\sim\times 3$ for our smallest systems, but at $M_*\ga 10^{11}M_\odot$ there is little evolution.
    \item Galaxy photometric projected sizes are in good agreement with observations for star-forming systems, but are too large for quenched systems particularly at lower masses.
    \item There is substantial evacuation of baryons from halos at group scales, with Local Group-sized objects retaining typically only a third of their cosmic baryon fraction.  Hot halo gas fractions show a rising trend with halo mass in good agreement with data.
    \item The black hole mass--stellar mass relation shows a population of quenched galaxies that agrees well with observations of bulge-dominated systems, while star-forming galaxies at a given $M_*$ have lower black hole masses.
    \item \simba\ predicts a $z=0$ dust mass function and dust-to-gas ratios in agreement with observations, with more massive star-forming galaxies having higher ratios.  At a given metallicity, galaxies that have higher SFR have higher dust-to-gas ratios. {\color{black} Roughly one-third of metals are locked in dust.}
\end{itemize}

These results demonstrate that \simba\ broadly reproduces a wide range of observed galaxy properties including their stellar content, star formation rates, gas content, metallicities, sizes, black hole properties, and dust properties.  Clearly there are many other more detailed constraints that would be tested, and in subsequent work we aim to examine these in more detail.  It is important to note that \simba\ also displays some aspects that are in conflict with observations.  It fails to produce as sharp a knee in the $z=0$ stellar mass function as is observed.  It produces low-mass quenched galaxy sizes that are larger than for star-forming systems, opposite to the observed trend in SDSS.  There are suggestions that \simba\ overproduces the stellar metallicities as well as the specific SFRs in low-mass present-day star-forming systems.  Finally, in many low-$z$ scaling relations (e.g. gas and metallicities vs. $M_*$) there is an abrupt break in the typical properties of galaxies above and below $M_*\approx 2\times 10^{10}M_\odot$, which qualitatively agrees observations but quantitatively may be too sharp. This rapid transition contrasts with the too-gradual turn-down in the GSMF, suggesting a tension in how \simba\ (and similar cosmological models) quench massive galaxies at low redshifts.  Despite these minor issues, \simba\ provides a state of the art platform for studying galaxy evolution across a range of masses, environments, and cosmic epochs, and promises to yield numerous new insights into the physical drivers of galaxy evolution over cosmic time.

\section*{Acknowledgements}

The authors acknowledge helpful discussions with Josh Borrow, Weiguang Cui, Shuiyao Huang, Katarina Kraljic, and Neal Katz.  We thank Philip Hopkins for making \gizmo\ public, and providing our group with early access. We thank Robert Thompson for developing {\sc Caesar}, and the {\sc yt} team for development and support of {\sc yt}.
RD acknowledges support from the Wolfson Research Merit Award program of the U.K. Royal Society.
DAA  acknowledges  support  by a  Flatiron  Fellowship.   The  Flatiron  Institute  is  supported by  the  Simons  Foundation. DN and QL were supported in part by NSF Award AST-1715206 and HST Theory Award 15043.0001.  
\textcolor{black}{
This work used the DiRAC@Durham facility managed by the Institute for Computational Cosmology on behalf of the STFC DiRAC HPC Facility. The equipment was funded by BEIS capital funding via STFC capital grants ST/P002293/1, ST/R002371/1 and ST/S002502/1, Durham University and STFC operations grant ST/R000832/1. DiRAC is part of the National e-Infrastructure.}

%%%%%%%%%%%%%%%%%%%%%%%%%%%%%%%%%%%%%%%%%%%%%%%%%%



\bibliographystyle{mnras}
\bibliography{simba} % if your bibtex file is called example.bib

%\begin{thebibliography}{99}
%\end{thebibliography}

%%%%%%%%%%%%%%%%%%%%%%%%%%%%%%%%%%%%%%%%%%%%%%%%%%

%%%%%%%%%%%%%%%%% APPENDICES %%%%%%%%%%%%%%%%%%%%%

%\appendix


% Don't change these lines
\bsp	% typesetting comment
\label{lastpage}
\end{document}